\documentclass[journal=jacsat,manuscript=article]{achemso}

\usepackage{chemformula} 
\usepackage[T1]{fontenc}

\usepackage[utf8]{inputenc}
\usepackage{amssymb,latexsym}
\usepackage{amsbsy} 
\usepackage{amsmath,amsthm}

\usepackage[shortlabels]{enumitem}
\usepackage{hyperref}

\usepackage{graphicx}
\usepackage{color}
\usepackage{caption}
\usepackage{subcaption}

\usepackage{tikz}
\usepackage{algorithmic}
\usepackage{wrapfig}

\usepackage{xr}
\usepackage{pdfpages}

\makeatletter
\newcommand*{\addFileDependency}[1]{
\typeout{(#1)}
\@addtofilelist{#1}
\IfFileExists{#1}{}{\typeout{No file #1.}}
}\makeatother

\newcommand*{\myexternaldocument}[1]{%
\externaldocument{#1}%
\addFileDependency{#1.tex}%
\addFileDependency{#1.aux}%
}

\myexternaldocument{supporting_information}

\theoremstyle{plain}

\theoremstyle{remark}

\theoremstyle{definition}

\newcommand \bra[1] {\langle {#1} |}
\newcommand \ket[1] {| {#1} \rangle}

\renewcommand{\epsilon}{\varepsilon}
\renewcommand{\kappa}{\varkappa}
\renewcommand{\phi}{\varphi}

\newcommand{\VC}{\mathcal{V}} 

\newcommand \commentout[1] {}

\author{Svala Sverrisdóttir}
\affiliation{Department of Mathematics, The University of California, Berkeley, CA 94720, USA}
\author{Fabian M. Faulstich}
\affiliation{Department of Mathematics, Rensselaer Polytechnic Institute, Troy, NY 12180, USA}
\email{faulsf@rpi.edu}

\title{Exploring Ground and Excited States via Single Reference Coupled-Cluster Theory and Algebraic Geometry}

\begin{document}

\begin{tocentry}
\begin{center}
\includegraphics[width = \textwidth]{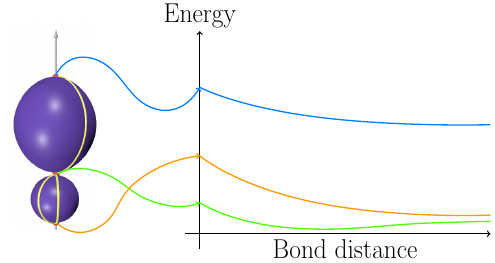}
\end{center}
\end{tocentry}

\begin{abstract}
The exploration of the root structure of coupled cluster equations holds both foundational and practical significance for computational quantum chemistry. This study provides insight into the intricate root structures of these non-linear equations at both the CCD and CCSD level of theory. We utilize computational techniques from algebraic geometry, specifically the monodromy and parametric homotopy continuation methods, to calculate the full solution set. We compare the computed CC roots against various established theoretical upper bounds, shedding light on the accuracy and efficiency of these bounds. We hereby focus on the dissociation processes of four-electron systems such as (H$_2$)$_2$ in both D$_{2{\rm h}}$ and D$_{\infty {\rm h}}$ configurations, H$_4$ symmetrically distorted on a circle, and lithium hydride. We moreover investigate the ability of single-reference coupled cluster solutions to approximate excited state energies. We find that multiple CC roots describe energies of excited states with high accuracy. Our investigations reveal that for systems like lithium hydride, CC not only provides high-accuracy approximations to several excited state energies but also to the states themselves. 
\end{abstract}

\section{Introduction}

Coupled cluster (CC) theory is a high-precision wave function method in computational quantum chemistry~\cite{bartlett2007coupled}. Its origins date back to 1958 when Coester proposed to use an exponential parametrization of the wave function~\cite{coester1958bound}, i.e.
\begin{equation}
\label{eq:ExponentialAnsatz}
\ket{\Psi}
=
e^{\hat T} \ket{\Phi_0},
\end{equation}
where $\hat T$ is the new unknown, and $\ket{\Phi_0}$ is a chosen reference Slater determinant (commonly the Hartree-Fock state). This parametrization was independently derived by Hubbard~\cite{hubbard1957description} and Hugenholtz \cite{hugenholtz1957perturbation} in 1957 as an alternative to summing many-body perturbation theory contributions order by order. A pivotal moment in the development of CC theory was {\v{C}}{\'i}{\v{z}}ek's seminal work in 1966~\cite{vcivzek1966correlation}. {\v{C}}{\'i}{\v{z}}ek presented the first derivation of the CC working equations together with corresponding simulations. This work established CC theory within the theoretical framework that we recognize today, including second quantization applied to many-fermion systems, normal ordering, contractions, Wick's theorem, normal-ordered Hamiltonians and Goldstone-style diagrammatic techniques. There exist several reviews of CC theory, summarizing different aspects of the method's success over the past decades, see e.g. K\"ummel~\cite{kummel1991origins}, {\v{C}}{\'i}{\v{z}}ek~\cite{vcivzek1991origins}, Bartlett~\cite{bartlett2005theory}, Paldus~\cite{paldus2005beginnings}, Arponen~\cite{arponen1991independent} and Bishop~\cite{bishop1991overview}. \\

Employing the exponential ansatz in Eq.~\eqref{eq:ExponentialAnsatz} yields the following reformulation of the stationary Schrödinger equation:
\begin{equation}
\label{eq:SEiffCC}
\left\lbrace
\begin{aligned}
\bra{\Phi_0} e^{-\hat T} \hat H e^{\hat T} \ket{\Phi_0}
&= E,\\
\bra{\Phi} e^{-\hat T} \hat H e^{\hat T}\ket{\Phi_0}
&= 0 &\forall \ket{\Phi}  \perp \ket{\Phi_0}.\\
\end{aligned}
\right.
\end{equation}
Employing the standard Galerkin projection, the latter condition yields the projected CC equations (vide infra), i.e.~a set of nonlinear algebraic equations. Typically, a single solution to this set of equations is computed using (quasi) Newton-type methods~\cite{helgaker2014molecular}. However, as a set of nonlinear algebraic equations, the CC equations possess multiple roots, raising the following natural questions: 
\begin{itemize}
    \item[i)] How many solutions to the CC equations exist?
    \item[ii)] Which solution describes the ground state best?
    \item[iii)] Do some of the solutions describe excited states?
\end{itemize}
Especially the last question is controversial, to say the least. On one side, it can be proven that as long a state is not strictly orthogonal to the chosen reference state, CC theory can -- in its untruncated form -- describe that state exactly. On the other side, excited states commonly have poor overlap with the Hartree-Fock ground state reference -- hindering the convergence of (quasi) Newton-type methods. Besides such theoretical arguments, it has been numerically shown that higher-order CC solutions can approximate excited state energies well while the CC solutions provide poor approximations to the actual states~\cite{piecuch2000computational}. Conventionally, excited states are computed in a post-single-reference CC manner. This is done by either constructing and diagonalizing the similarity-transformed Hamiltonian in equation-of-motion CC~\cite{rowe1968equations,emrich1981extension,sekino1984linear,geertsen1989equation,stanton1993equation,comeau1993equation,watts1994inclusion}, or by examining the poles of the linear-response function in linear-response CC~\cite{sekino1984linear,monkhorst1977calculation,dalgaard1983some,koch1990coupled}. In either case, the excited states provided by these methods are inherently biased toward the lowest eigenstate upon which they are built. As a direct consequence, the accuracy of such approaches is known to be quite poor for certain classes of eigenstates that are energetically higher than the ground state~\cite{watson2012excited,shu2017doubly,barca2018excitation,loos2019reference,ravi2022intermediate,do2023classification}.
An alternative approach, which aligns more closely with the ideas discussed in this work and offers greater accuracy for doubly excited states compared to EOM-CC\cite{lee2019excited}, is known as $\Delta$CC. This method directly employs the CC wavefunction parametrization for excited states\cite{meissner1993multiple,jankowski1994applicability,jankowski1994multiple,jankowski1995multiple,jankowski1999physical1,jankowski1999physical2,jankowski1999physical3,podeszwa2002multiple,podeszwa2003multiple,mayhall2010multiple}.

Besides the fundamental nature of questions i)--iii), the existence of multiple roots to the CC equations can present challenges for the practical implementations of the method. For example, the roots may be difficult to adequately converge to~\cite{vcivzek1966correlation,piecuch1994solving,mihalka2023exploring,shavitt2009many,lee2019excited}, and the convergence properties of the iterative solution methods can strongly depend on the employed initial guess~\cite{szakacs2008stability,adams1981symmetry}. See also Ref.~\citenum{faulstich2024recent} for a mathematical exposition of this problem and Ref.~\citenum{faulstich2024augmented} for a potential remedy.

In this work, we elaborate on recent advances coming from algebraic geometry that provide new ideas on how to bound the number of roots to the CC equations. These new and significantly improved bounds can then be used to compute all roots to the CC equations, answering the above-posed questions i) -- iii). In particular, we demonstrate that higher-order solutions of the CC equations can effectively approximate not only the energies of excited states but also the eigenstates themselves.

\subsection{Previous works and perspective}

We highlight that the inaugural investigation of the root structure of CC methods dates back to 1978 when Zivkovic and Monkhorst scrutinized the singularities and multifarious solutions in single-reference CC equations~\cite{vzivkovic1978analytic}. Building upon this work, Paldus and colleagues subsequently conducted mathematical and numerical analyses during the early 1990s, focusing on the manifold of solutions of both single-reference and state-universal multi-reference CC equations, while also elucidating their singularities and analytical characteristics~\cite{paldus1993application,piecuch1990coupled}. In 1998, Kowalski and Jankowski revived homotopy methods in connection with the single-reference CC theory and used them to solve a CCD system~\cite{kowalski1998towards}. Kowalski with Piecuch then extended the application of homotopy methods to the CCSD, CCSDT, and CCSDTQ~\cite{piecuch2000computational} for a 4-electron system described in a minimal basis set. From these investigations, they derived the $\beta$-nested equations and proved the {\it Fundamental Theorem of the $\beta$-NE Formalism}~\cite{piecuch2000computational}, which enabled an explanation of curves connecting multiple solutions of the various CC polynomial systems, i.e.~from CCSD to CCSDT, CCSDT to CCSDTQ, etc. This {\it Kowalski-Piecuch homotopy} was recently analyzed in-depth using topological degree theory~\cite{csirik2023disc,csirik2023coupled}. Moreover, Kowalski with Piecuch used homotopy methods to obtain all solutions of the generalized Bloch equation~\cite{kowalski2000complete} and to compute all solutions to the multireference coupled-cluster equations within the framework of the state-universal coupled-cluster formalism\cite{kowalski2000complete2}. These works extend the preceding works by Kowalski and Jankowski\cite{kowalski1998full,jankowski1999physical1,jankowski1999physical2,jankowski1999physical3,jankowski1999physical4}. 
It is noteworthy that these fundamental results of CC theory have spurred the development of the method of moment CC approach~\cite{kowalski2000method,piecuch2002recent,piecuch2004method,shen2012biorthogonal}, which in turn has led to the highly efficient completely renormalized CC approach~\cite{kowalski2000renormalized,piecuch2006single,piecuch2005renormalized}.
This emphasizes the pivotal role of this research trajectory in driving methodological innovations within CC theory. A perspective on the recent use of homotopy methods in the context of quantum chemistry can be found in Ref.~\citenum{faulstich2023homotopy}. 

A more algebraic and mathematically driven approach to this problem has recently gained momentum. Initialized by some of the authors, a first algebraic perspective was introduced in Ref.~\citenum{faulstich2024coupled}. This was followed by an in-depth algebraic investigation leading to the introduction of the CC truncation varieties~\cite{faulstich2023algebraic} -- which are the subject of this work. This work moreover contained open mathematical questions and conjectures, one of which was proven shortly after~\cite{borovik2023coupled}. More generally, these works are part of a presently increasing interest in the mathematical understanding of CC theory, driven by the applied mathematics community. In this context, Schneider performed the first local analysis of CC theory in 2009 employing a functional analytic framework~\cite{schneider2009analysis}. Subsequently, this approach was refined~\cite{rohwedder2013continuous,rohwedder2013error} and applied to different CC methods~\cite{laestadius2018analysis,laestadius2019coupled,faulstich2019analysis}. A more versatile framework for analyzing general CC variants using topological degree theory was later proposed by Csirik and Laestadius~\cite{csirik2023disc,csirik2023coupled}.   
The most recent numerical analysis results regarding single reference CC were established by Hassan, Maday, and Wang~\cite{hassan2023analysis,hassan2023analysis2}.

\section{Coupled-cluster theory}
\label{sec:CC-theory}

The starting point of CC theory is the exponential ansatz in Eq.~\eqref{eq:ExponentialAnsatz}, leading to the CC energy expression
\begin{equation}
E
=
\bra{\Phi_0} e^{-\hat T} \hat H e^{\hat T} \ket{\Phi_0},
\end{equation}
where the {\it cluster operator} $\hat T$ is determined through the CC equations
\begin{equation}
\label{eq:projections}
0 = \bra{\Phi} e^{-\hat T} \hat H e^{\hat T} \ket{\Phi_0},\qquad \forall \ket{\Phi} \perp \ket{\Phi_0}.
\end{equation}

The cluster operator is a linear combination of elementary particle-hole excitation operators\cite{vcivzek1966correlation}, i.e. 
\begin{equation}
\label{cluster matrices}
\hat T({\bf t}) = \sum_{\mu > 0 } t_\mu \hat X_\mu,
\end{equation}
where we highlight the dependence of $\hat T$ on its expansion coefficients by writing $\hat T(\bf t)$.
It is therefore convenient to label the elementary particle-hole excitation operators using multi-indices $\mu>0$ that clarify the projections and wedge products involved~\cite{helgaker2014molecular, schneider2009analysis}. Therefore, acting on the reference state $\ket{\Phi_0}$, the elementary particle-hole excitation operators define the Hilbert space of functions that are $L^2$-orthogonal to $\ket{\Phi_0}$.
We henceforth set 
$
\VC := \{\ket{\Phi_0} \}^\perp 
$
and note that $\hat T({\bf t}) \ket{ \Phi_0 }\in \VC$ for all CC amplitudes $\mathbf{t}$. 
Hence, we can express Eq.~\eqref{eq:projections} in the more common form as a set of projective equations, i.e.  
\begin{equation}
\label{eq:OrthogonalCond}
0 = \langle  \Phi_\mu | e^{-\hat T({\bf t})} \hat H e^{\hat T({\bf t})} | \Phi_0\rangle, \quad \forall \mu>0,
\end{equation}
where $\ket{\Phi_\mu} = \hat X_\mu \ket{\Phi_0}$ are the excited Slater determinants spanning $\VC$.
In practice, the expansion in Eq.~\eqref{cluster matrices} is commonly limited to e.g.~only containing one and two particle-hole excitations leading to CCSD. In this case, the projective equations \eqref{eq:OrthogonalCond} are also restricted to the corresponding set of excited Slater determinants, yielding a square system of polynomial equations that we seek to solve. Hence, a key object in CC theory is the set 
\begin{equation}\label{eq:CC-variety}
\mathcal{S}
=
\{
{\bf t} \in \mathbb{F}^K ~|~ \langle  \Phi_\mu | e^{-\hat T({\bf t})} \hat H e^{\hat T({\bf t})} | \Phi_0\rangle = 0, \quad \forall \mu >0
\},
\end{equation}
where $\mathbb{F}$ is the considered number field (either $\mathbb{R}$ or $\mathbb{C}$) and $K$ is the ``system size'' (determined by the number of correlated electrons, the size of one-particle basis functions as well as further selection rules). Note that $\mathbb{F}$ can be $\mathbb{R}$ or $\mathbb{C}$ depending on whether we are seeking real- or complex-valued CC amplitudes. We emphasize that although the CC polynomial coefficients are real, the roots of the polynomial system may not be. Mathematically, the theory describing complex-valued solutions is simpler and more complete than the theory describing real-valued solutions. Therefore, we shall consider $\mathbb{F} = \mathbb{C}$ for mathematical considerations, however, solving the truncated CC equations over $\mathbb{F} = \mathbb{C}$ may yield unwanted complex valued energies. 

\begin{figure}[h!]
    \begin{center}
    \includegraphics[width = 0.35\textwidth]{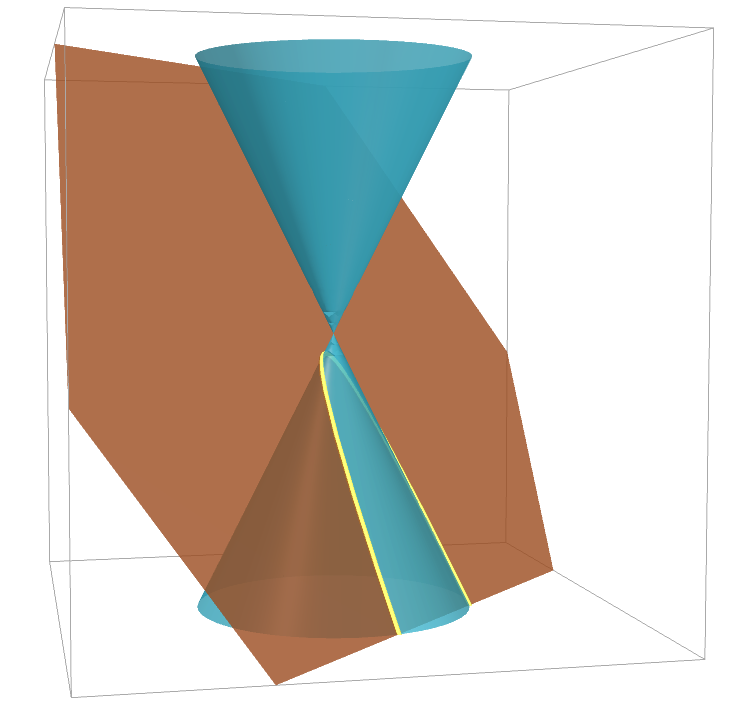}
    \end{center}
    \caption{Root structure corresponding to Eq.~\eqref{eq:simplevarity}. The cone describes the roots of $p_1$, and the plane describes the roots of $p_2$. The simultaneous roots are described by the yellow curve that defines the solution variety of the multivariate polynomial system.}
    \label{fig:simpleVariety}
\end{figure}

Mathematically, the set $\mathcal{S}$ is an (algebraic) variety. Since the concept of varieties may be unconventional within the computational chemistry community, we provide an illustrative example.
A variety is a mathematical object defined by the zeros of a set of polynomial equations. Consider the multivariate polynomial system 
\begin{equation}
\label{eq:simplevarity}
\left\lbrace
\begin{aligned}
p_1(x,y,z) &= x^{2}+y^{2}-z^{2},\\
p_2(x,y,z) &= \frac{x + y}{\sqrt{2}}+z-1.
\end{aligned}
\right.
\end{equation}
The zeros of $p_1$ form a cone, whereas the zeros of $p_2$ form a plane; the set of simultaneous roots defines a variety that corresponds to a parabolic curve, see Fig.~\ref{fig:simpleVariety}. As a curve, this variety is one-dimensional, embedded in the ambient space which is three-dimensional in this case. A central quality for this work is the \textit{degree} of a variety which is defined to be the maximal number of intersection points of the variety with a general linear space of complementary dimension. By requiring the linear space to be general we ensure the intersection is always finite. For the variety in Fig.~\ref{fig:simpleVariety} we see that intersecting the yellow curve with a plane (a linear space of complimentary dimension) yields at most two intersection points, hence, its degree is two. 
We add that over the complex field, the condition of ``maximal number of intersection points''  is redundant for the definition of degree since the number of intersection points is the same for generic complex spaces of complementary dimension. 

We conclude this section by noting that in the case of untruncated CC, amplitudes that solve the full CC equations describe eigenstates of the Hamiltonian. Therefore, the cardinality of $\mathcal{S}$ is exactly the number of eigenstates of $\hat{H}$ that are intermediately normalized and the number of roots to the CC equations is bounded from above by the number of Slater determinants. However, when truncations are imposed, the cardinality of $\mathcal{S}$ increases, and can ultimately lead to more solutions than provided by full diagonalization of the system. Consequently, it becomes less clear what the different elements in $\mathcal{S}$ describe. Understanding the variety $\mathcal{S}$ and characterizing (some of) its elements are the subject of this manuscript. 

\section{Truncation varieties}

In CC theory one encounters various algebraic equations that define different varieties. The {\it truncation varieties}~\cite{faulstich2023algebraic} arise from the exponential parametrization, i.e.
\begin{equation}
{\rm exp}:\mathbb{V} \to \mathcal{H},~\mathbf{t} \mapsto e^{\hat{T}(\mathbf{t})} |\Phi_0\rangle,
\end{equation}
where $\mathbb{V}$ is the CC amplitude space and $\mathcal{H}$ denotes the space of intermediately normalized wave functions. In the untruncated CC method, the exponential map is bijective. However, when truncations are introduced, this property no longer holds. Truncations correspond to restrictions of the exponential map to a subspace, e.g. for CCSD the exponential map is restricted to $\mathbb{V}_{\{1,2\}} \subset \mathbb{V}$ consisting of cluster amplitudes with only one and two particle-hole excitations. The corresponding truncation variety, $V_{\{1,2\}}$ is then defined as the smallest variety containing ${\rm exp}(\mathbb{V}_{\{1,2\}})$. This construction of the truncation variety can be generalized to any CC truncation: Let $\sigma$ denote the CC truncation rank, i.e.~$\sigma \subset [\![N]\!] = \{1, \dots, N\}$, where $N$ denotes the number of electrons in the system. Then, the smallest variety containing ${\rm exp}(\mathbb{V}_\sigma)$ is the $\sigma$-truncation variety denoted $V_\sigma$.  

In this article, we employ the unlinked CC equations~\cite{}, which correspond to 
\begin{equation}
\label{eq:newCCeqs}
[\hat H | \Phi \rangle]_\sigma = E [| \Phi \rangle]_\sigma, \quad |\Phi\rangle \in V_\sigma,
\end{equation}
where $[\,\cdot\,]_\sigma$ is the projection onto coordinates with excitation level in $\sigma \cup \{0\}$, i.e.
\begin{equation}
\label{eq:CCLinear}
[|\Psi\rangle ]_\sigma = \langle \Phi_\mu |\Psi\rangle \quad {\rm for}~|\mu|\in\sigma  \cup \{0\}.  
\end{equation}
This formulation is equivalent to the projective CC equations in Eq.~\eqref{eq:OrthogonalCond} for CCD, as well as for rank complete CC truncations, i.e.~CCSD, CCSDT, CCSDTQ etc., see e.g.~Ref.~\citenum{coester1960short} for the original proof of this statement or Theorem 5.11 in Ref.~\citenum{faulstich2023algebraic}.
The number of roots for these CC variants is bounded by the number of solutions to Eq.~\eqref{eq:newCCeqs} for a generic Hamiltonian matrix $\hat H$. We denote this number the \textit{CC degree} of the corresponding truncation variety.
The CC degree serves as a complexity measure for solving the CC equations and Bézout's theorem provides the upper bound
\begin{equation}
\label{eq:VarietyBound}
\mathrm{CCdeg(\sigma)} \leq (\dim(V_\sigma) + 1) \deg (V_\sigma),    
\end{equation}
where $\dim(V_\sigma)$ is the {\it dimension} of the truncation variety and $\deg (V_\sigma)$ is its {\it degree}. The dimension of the truncation variety is well understood. The reason is that cluster analysis allows us to bijectively map cluster amplitudes to CI-coefficients using polynomial equations, see e.g.~Refs.~\citenum{helgaker2014molecular,crawford2007introduction,shavitt2009many} or Proposition 2.4 in Ref.~\citenum{faulstich2023algebraic} for a mathematical proof. Then, since $V_\sigma$ arises as a deformation of the linear space $\mathbb{V}_\sigma$ under a polynomial with polynomial inverse, the dimension of $V_\sigma$ is equal to the dimension of $\mathbb{V}_\sigma$. For example in the case of CCSD, we obtain
\begin{equation}
\dim (V_\sigma) =\dim (\mathbb{V}_\sigma) = n_{\rm virt}n_{\rm occ} + \binom{n_{\rm virt}}{2} \binom{n_{\rm occ}}{2}.
\end{equation}
The degree of ${V}_\sigma$ on the other hand is not as easily computable and requires a more profound understanding of ${V}_\sigma$. For special cases, however, explicit formulas for the degree have been established; for example, for CCS, the truncation variety is the Grassmanian (see Theorem 3.5 in Ref.~\citenum{faulstich2023algebraic}) which is mathematically well understood. In other cases we rely either on symbolic methods, provided by e.g.~\texttt{Macaulay2} \cite{M2}, or on numerical methods, accessible in e.g.~\texttt{HomotopyContinuation.jl} \cite{breiding2018homotopycontinuation}, to compute the degree. 

Despite the complexity of calculating $\deg (V_\sigma)$ the bound given by Eq.~\eqref{eq:VarietyBound} provides significant insight into bounding the CC degree, since it is less complex to calculate $\deg (V_\sigma)$ than to calculate the CC degree directly. Moreover, this upper bound is notably better than the previously known upper bounds obtained by applying B\'ezout's theorem directly to the projective CC equations, see Section 6 in Ref.~\citenum{faulstich2023algebraic}. 

\section{Computational methods}

In this section, we outline the computational procedure employed to compute the CC degree and the CC roots for a given CC truncation rank $\sigma \subset [\![ N ]\!]$. We begin by introducing the CC family, a {\it parametric polynomial family} for Eq.~\eqref{eq:newCCeqs}, i.e.
\begin{equation}
\mathcal{F}_{\rm CC}(\sigma) 
= 
\{ [ (\hat G - E \cdot I_K) e^{\hat T(\mathbf{t})} |\Phi_0\rangle]_\sigma ~\big|~ \hat G \in \mathbb{F}^{K\times K} \},   
\end{equation}
where the energy $E\in\mathbb{C}$ and cluster amplitudes $\mathbf{t} \in \mathbb{V}_\sigma$ are the variables, and the entries of the matrix $\hat G$ are the parameters. We will ultimately set $\hat G = \hat H$ to be the Hamiltonian of interest, but $\mathcal{F}_{\rm CC}(\sigma)$ is a much broader family of CC-like systems with $\hat G \in \mathbb{F}^{K \times K}$ an arbitrary matrix. Note that the matrix $\hat G$ can be parameterized including further quantum mechanical structures, however, experiments have shown that this does not result in significantly fewer roots while yielding slower calculations due to the inclusions of further constraints.
The {\it parameter continuation theorem} provides an upper bound to the number of roots of any polynomial system $F_{\hat G} \in \mathcal{F}_{\rm CC}$. Moreover, this bound is tight and equality is obtained for all polynomial systems with parameters in a  
certain dense subspace of $\mathbb{F}^{K\times K}$.
We subsequently assume the validity of the parameter continuation theorem and refer the interested reader to Section 3 in Ref.~\citenum{breiding2024metric}. The upper bound provided by the parameter continuation theorem yields the CC degree mentioned in the previous section. 

The numerical procedure used in this work consists of two steps: First, we solve a general system of equations from the CC family, $\mathcal{F}_{\rm CC}$, using {\it monodromy methods}. Second, we use {\it parameter homotopy continuation} to track the solutions from this solved but generic system of equations to solutions of the CC equations describing a specific chemical system of interest. Both steps are subsequently outlined in detail. 

We emphasize that, at their core, both methods use a \textit{homotopy} to track new solutions from known solutions~\cite{bates2023numerical,garcia1979finding,morgan2009solving,sommese2005numerical}. More precisely, we define a homotopy for a (piecewise smooth) path $\gamma: [0,1] \to \mathbb{F}^{K \times K}$ in the parameter space via
\begin{equation}
H(E, \mathbf{t}, \lambda) = F_{\gamma(\lambda)}(E, \mathbf{t}),
\end{equation}
where $H(E, \mathbf{t}, \lambda)$ describes the continuous deformation of a start system $G(E,\mathbf{t}) = H(E, \mathbf{t}, 1)$ to a target system $F(E, \mathbf{t}) = H(E, \mathbf{t}, 0)$.
The individual solution paths to the system $H(E, \mathbf{t}, \lambda)$ are then tracked from $G(E,\mathbf{t})$ to $F(E, \mathbf{t})$ as $\lambda \to 0$. This is accomplished via an ordinary differential equation, known as the Davidenko differential equation~\cite{davidenko1953new,davidenko1953approximate}, i.e.
\begin{equation}
\label{eq:Davidenko}
\frac{\partial}{\partial \mathbf{x}} H(\mathbf{x}, \lambda)\left(\frac{\mathrm{d}}{\mathrm{d} \lambda} \mathbf{x}(\lambda)\right)+\frac{\partial}{\partial \lambda} H(\mathbf{x}, \lambda)=0,
\end{equation}
initialized by a root $\mathbf{x}$ of the start system, i.e.~$G(\mathbf{x}) = 0$ where we set $\mathbf{x} =(E,\mathbf{t}) \in \mathbb{C}\times \mathbb{V}_{\sigma}$.

\subsection{Monodromy}
\label{sec:Monodromy}

The monodromy solver is a numerical technique specifically designed for solving generic polynomial systems within a parameterized family. We emphasize that the monodromy solver requires the system to be generic, meaning it is not suitable for solving a specific set of CC equations directly. However, it can be employed to establish a start system $F_{\hat{G}}\in\mathcal{F}_{\rm CC}$ within the CC family. This start system is then used by the parameter homotopy continuation method to solve a (physical) target system of interest. Note that the monodromy method is required only once for a particular system configuration, i.e.~the number of electrons, the number of orbitals, and the CC truncation level $\sigma\in[\![N ]\!]$.

We start with a root $\mathbf{x}\in \mathbb{C}\times \mathbb{V}_{\sigma}$ of a generic system $F_{\hat G} \in \mathcal{F}_{\rm CC}$ in the CC family and a loop $\gamma: [0,1] \to \mathbb{F}^{K \times K}$ in the parameter space, based at the generic matrix $\hat G$, i.e.~$\gamma(0) = \gamma(1) = \hat G$. Moreover, we let $\mathbf{x}(\lambda)$ be the solution path of $H(E, \mathbf{t}, \lambda) = 0$ along $\gamma$, starting at $\mathbf{x}(1) = \mathbf{x}$. Since $\gamma$ is a loop, the start system in this homotopy is equal to the end system. In particular, the solution path ends again at a root of $F_{\hat G}$. The monodromy solver exploits the fact that the root $\mathbf{x}(0)$ need not necessarily be equal to the root $\mathbf{x}(1)$. In fact, let $(\mathbf{x}_1, \dots,\mathbf{x}_M)$ be all roots to $F_{\hat G}$, then the roots $(\mathbf{x}_1(0), \dots,\mathbf{x}_M(0))$ define a mere permutation the system's roots. 

\begin{figure}[h!]
    \begin{center}
    \includegraphics[width = 0.75\textwidth]{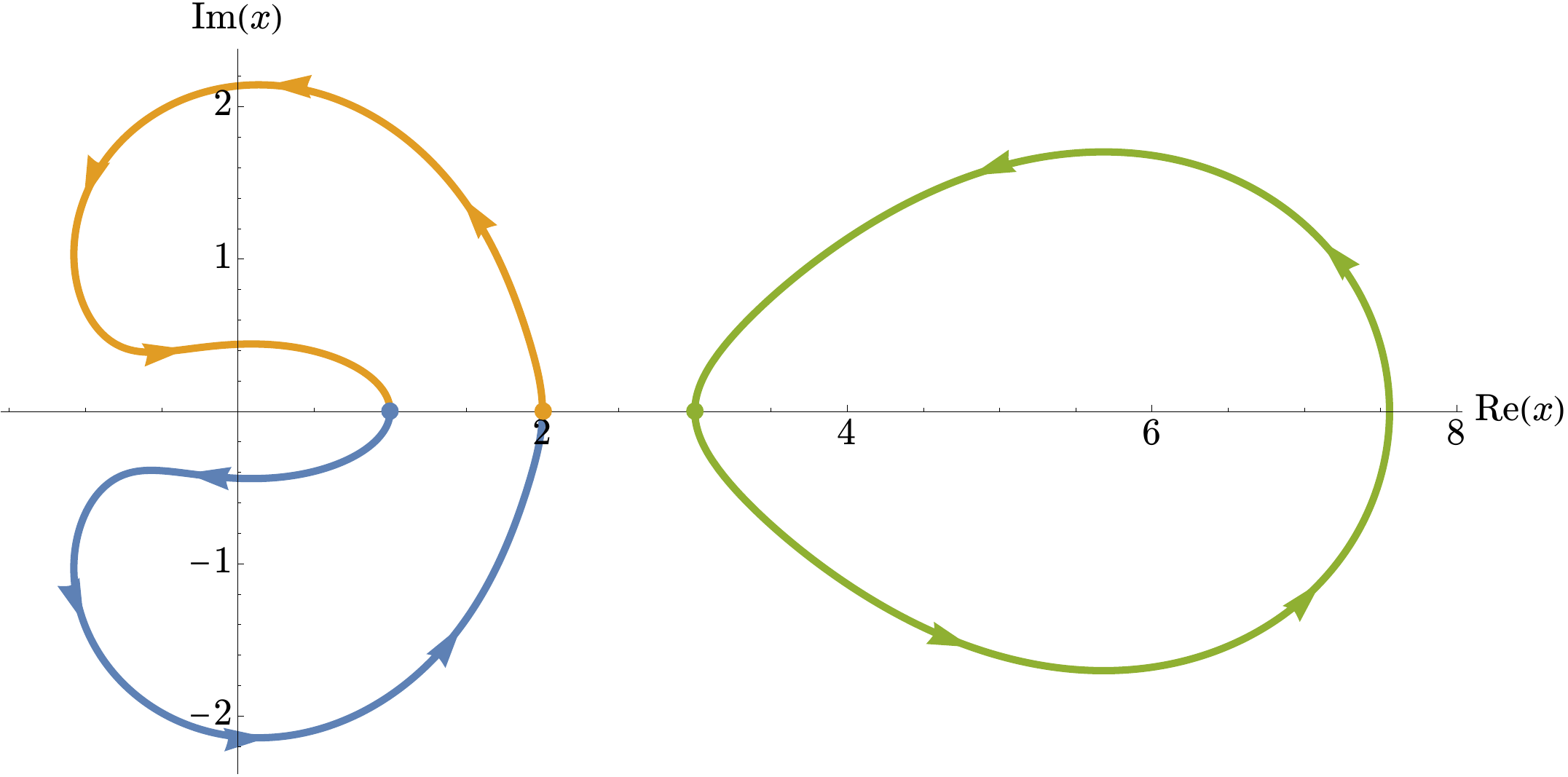}
    \end{center}
    \caption{The solution paths of the roots of $x^3 - 6x^2 + 11zx - 6$ for $z(\lambda) = \gamma(\lambda)$ along the path $\gamma:[0,1]\to\mathbb{C}, t \mapsto e^{i2\pi t}$.}
    \label{fig:monodromy}
\end{figure}

We exemplify this procedure with the simple parametric polynomial family:
\begin{equation}
\label{eq:MonoExample}
\mathcal{F} = \{ x^3 - 6\,x^2 + 11\,z\,x - 6 ~\big|~ z \in \mathbb{C}\},
\end{equation}
where $x\in \mathbb{C}$ is the variable and $z\in\mathbb{C}$ is the parameter. For $z = 1$, the corresponding polynomial 
\begin{equation}
\label{eq:MonoExample1}
x^3 - 6\,x^2 + 11\,x - 6
=(x-1)(x-2)(x-3)
\end{equation}
has the roots $x_1 = 1, x_2 = 2, x_3 = 3$. The monodromy solver then tracks these roots along the path $\gamma:[0,1]\to\mathbb{C}, t \mapsto e^{i2\pi t}$. Note that $\gamma$ is a loop based at $1$. Indeed, this procedure permutes the roots, see Figure~\ref{fig:monodromy}. 

This approach can be effectively utilized to find all solutions to a generic system of equations. For instance, consider a scenario where the only known solution for the polynomial in Eq.~\eqref{eq:MonoExample1} is $x_2 = 2$. By tracking this solution along $\gamma$, we can determine another solution, $x_1 = 1$. When we extend this tracking across sufficiently many loops, the method reveals all possible roots of the system. 

This process is central to the monodromy solver. The solver begins with a single known solution and numerically tracks the solution through a loop within the parameter space to find a new solution. The new known solution can now be tracked through the same loop. This procedure is continued until an already-known solution is found. Then we generate a new loop and repeat this for all known solutions.

To terminate this procedure a stopping criterion for the monodromy solver has to be chosen. As described above the loops in the parameter space define a permutation of the solution set. Therefore the probability of obtaining all solutions iterating through a fixed number of loops is equal to the probability that a fixed number of random permutations in the symmetry group of the solution set $R = \{\mathbf{x}_1, \dots, \mathbf{x}_M\}$, denoted by $S_R$, generate a \textit{transitive subgroup}. 
Recall that a transitive subgroup in $S_R$ is a subgroup such that we can permute any element in $R$ to another by a permutation in that subgroup.
Dixon\cite{babai1989probability} shows the probability that an ordered pair of random permutations in $S_R$ generates a transitive subgroup is 
\begin{equation}
1 - |R|^{-1} + O(|R|^{-2}).
\end{equation}
For a large solution set, stopping after two iterations can give all solutions with a high probability. However, it is more common to stop after finding no new solutions over a predetermined number of loops. In simulations presented here, we use the monodromy solver in \texttt{HomotopyContinuation.jl} and their default stopping criterion that stops the solver after $5$ loops with no new solutions.

The parameters in the CC family $\mathcal{F}_{\rm CC}$ are linear, hence, we can find a generic start system along with a root by choosing a pair $(E, \mathbf{t}) \in \mathbb{C} \times \mathbb{V}_\sigma$ at random, and then compute one generic solution $\hat G_0$ fulfilling
$$
[(\hat G_0 - E \cdot I_K) e^{\hat T(\mathbf{t})}]_\sigma = 0.
$$
We then use the monodromy solver to compute all roots to $F_{\hat G_0}$.

\subsection{Parametric homotopy continuation}

Applying the parametric homotopy continuation method to find the roots of the CC equations corresponding to a system of interest is now straightforward. The goal is to solve
\begin{equation}
F_{\hat H}(E,{\mathbf t})
= [(\hat H - E \cdot I_K) e^{\hat T(\mathbf{t})} ]_\sigma =0 
\end{equation}
for a given Hamiltonian $\hat H$, arising from a physical electronic system. By construction, both $\hat H$ and $\hat G_0$ are in the parameter space of $\mathcal{F}_{\rm CC}$. In particular, we can define a path $\gamma$ in the parameter space that continuously transforms $\hat G_0$ into $\hat H$. Since we have computed all roots to the system $F_{\hat G_0}$, the Davidenko differential equation~\eqref{eq:Davidenko} allows the tracking of solutions of $F_{\hat G_0}$ to $F_{\hat H}$ using e.g. Newton's method.

\begin{figure}[h!]
    \centering
    \includegraphics[width = 0.45 \textwidth]{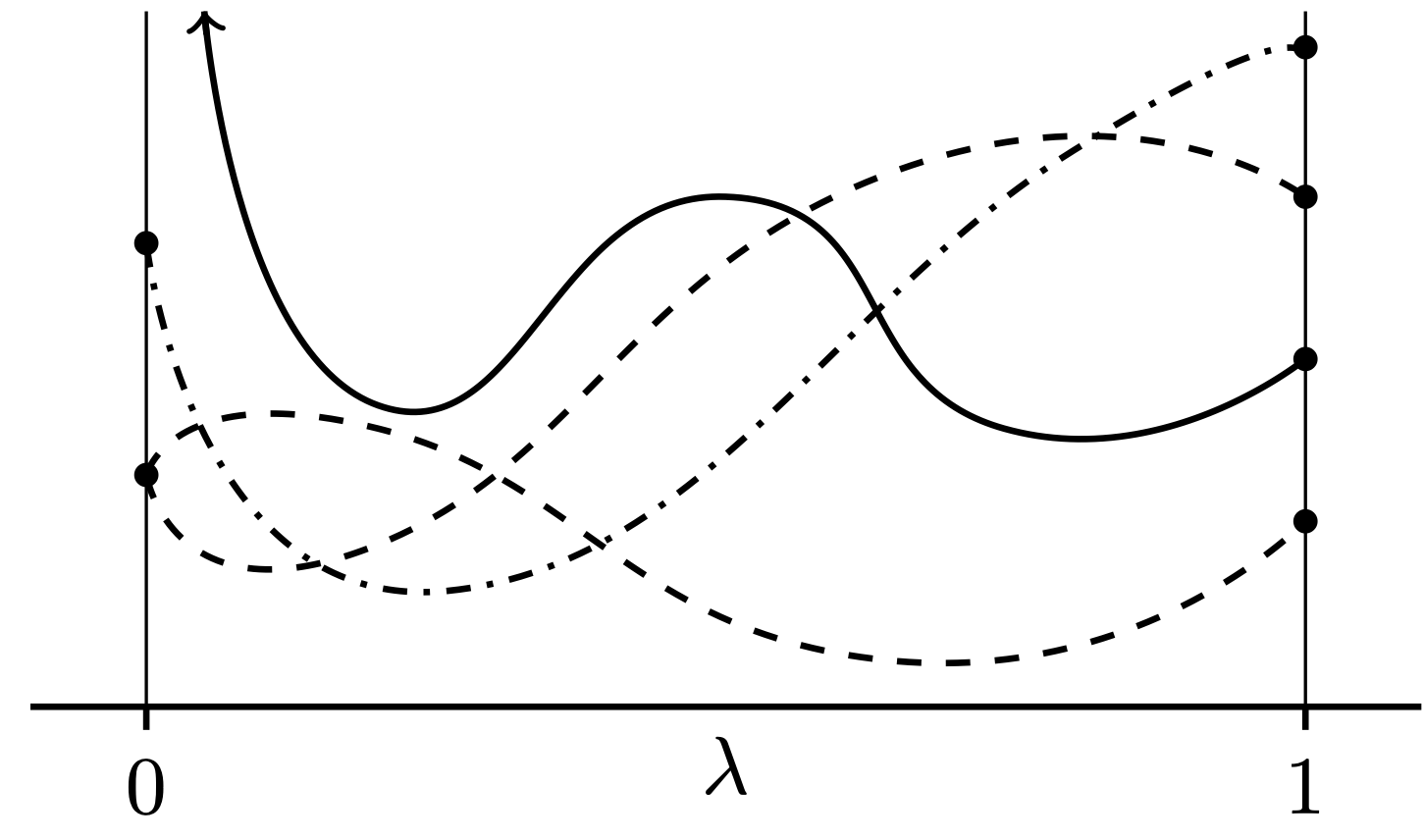}
    \caption{Sketch of possible homotopy paths\cite{faulstich2023homotopy}. The solid line shows a path with no finite limit as $\lambda \to 0$, the dashed lines have the same limit of a singular root, and the dotted-dashed line has a unique limit of a nonsingular root.}
    \label{fig:homotopies}
\end{figure}

As $\lambda \to 0$, we may encounter different scenarios when getting close to $\lambda=0$. Here, the {\it endgame}~\cite{morgan1992computing} should be employed to determine the target solutions and their ``type''. We illustrate such different scenarios in Figure~\ref{fig:homotopies}.

We conclude this section by highlighting the progress that has been made within homotopy continuation based software development exploiting parallel implementations, which can significantly extend their application areas in the coming years. In particular, we here want to emphasize PHCpack~\cite{verschelde1999algorithm}, Bertini,~\cite{bates2006bertini} HOM4PS,~\cite{chen2014hom4ps,lee2008hom4ps} NAG4M2~\cite{bates2023numerical} and 
\texttt{HomotopyContinuation.jl}.~\cite{breiding2018homotopycontinuation}

\section{Numerical results}

In this section we investigate the root structure and potential energy curves (PECs) for ground and excited states at the level of CCD and CCSD. We report PECs corresponding to different CC solutions obtained by the above-outlined computational procedure. The obtained energies and corresponding eigenstate approximations are then compared to exact eigenpairs of the respective Hamiltonian. Besides the computation and analysis of the CC states, we compare different bounds to the number of CC roots with the computed number of CC solutions along different bond stretching processes. We restrict our investigations to four electron systems, such as (H$_2$)$_2$ in D$_{2 {\rm h}}$ and D$_{\infty {\rm h}}$ configurations, H$_4$ symmetrically disturbed on a circle~\cite{van2000benchmark}, and LiH. The reported computations use a minimal basis description using the STO6G basis set and all considered systems at half-filling, i.e.~four electrons in eight spin orbitals. The one-body and two-body integrals for the considered systems are obtained from the Python-based Simulations of Chemistry Framework (\texttt{PySCF})~\cite{sun2020recent,sun2018pyscf,sun2015libcint}. For the algebraic simulations, we used the monodromy and parametric homotopy continuation methods in \texttt{HomotopyContinuation.jl}. \cite{breiding2018homotopycontinuation} 

\subsection{CC bounds for generic four electron systems at half-filling}

We emphasize that the theoretical bounds hold for generic Hamiltonians, and therefore only depend on the number of electrons, the number of spin orbitals, and the CC truncation level.

\paragraph{Coupled cluster doubles:}
The roughest bound is obtained by applying B\'ezout's theorem directly to the CCD equations. Note that for four electron systems, the CCD equations are quadratic. Moreover there are ${4\choose 2}^2 = 36$ doubly excited Slater determinants. Hence, the B\'ezout bound directly applied to the CCD equations yields $2^{36}$ as an upper bound for the number of roots.
The second bound we compare with is the bound in Eq.~\eqref{eq:VarietyBound}. There are 36 CCD amplitudes, hence, ${\rm dim}(V_{\{2\}}) = 36$. Moreover we compute the degree of $V_{\{2\}}$ using \texttt{Macaulay2} yielding ${\rm deg}(V_{\{2\}}) = 2$. Overall, this yields $74$ as an upper bound.
For a generic Hamiltonian, we can moreover compute the exact number of roots to the CCD equations using the monodromy method which yields CCdeg$_{4,8}(\{2\})=73$. We verify that these are all the solutions by running the functions \texttt{certify} -- checks that we have distinct solutions -- and $\mathtt{verify\_solution\_completeness}$ -- uses a trace test to check we have all solutions -- in \texttt{HomotopyContinuation.jl}. 
This quantifies the improvement provided by the bound in Eq.~\eqref{eq:VarietyBound} over the direct application of B\'ezout's theorem. 

\paragraph{Coupled cluster singles and doubles:}
When considering CCSD instead of CCD these numbers dramatically increase. First, we note that the B\'ezout bound directly applied to the CCSD equation yields the upper bound $3^{n_s}4^{n_d}$, where for four electrons in eight spin orbitals $n_s=16$ and $n_d=36$. This leads to approximately $2\cdot 10^{29}$ solutions. Second, we note that ${\rm dim}(V_{\{1,2\}}) = 52$ and numerical calculations yield ${\rm deg}(V_{\{1,2\}})= 442066$, hence, the bound provided by Eq.~\eqref{eq:VarietyBound} yields approximately $2 \cdot 10^7$ solutions. Numerical calculations show that CCdeg$_{4,8}(\{1,2\}) \approx 16\,952\,996$. Due to the size of the system we stopped the monodromy solver after only two loops, while it was still finding new solutions, so this number is an underestimation of the actual CC degree. 
This clearly illustrates the improvement provided by the bound in Eq.~\eqref{eq:VarietyBound} over the direct application of B\'ezout's theorem. \\

To connect these theoretical bounds to practical systems, we subsequently calculate the number of CC solutions that result in real-valued energies, as well as those that accurately predict excited state energies at different bond lengths for the considered systems. Additionally, we report the number of FCI solutions. This comparison allows us to assess the accuracy of the bounds established for generic Hamiltonians.

\subsection{Dissociation of lithium hydride -- CCD}

\begin{figure}[h!]
    \centering
    \begin{subfigure}[t]{0.45\textwidth}
    \includegraphics[width =\textwidth]{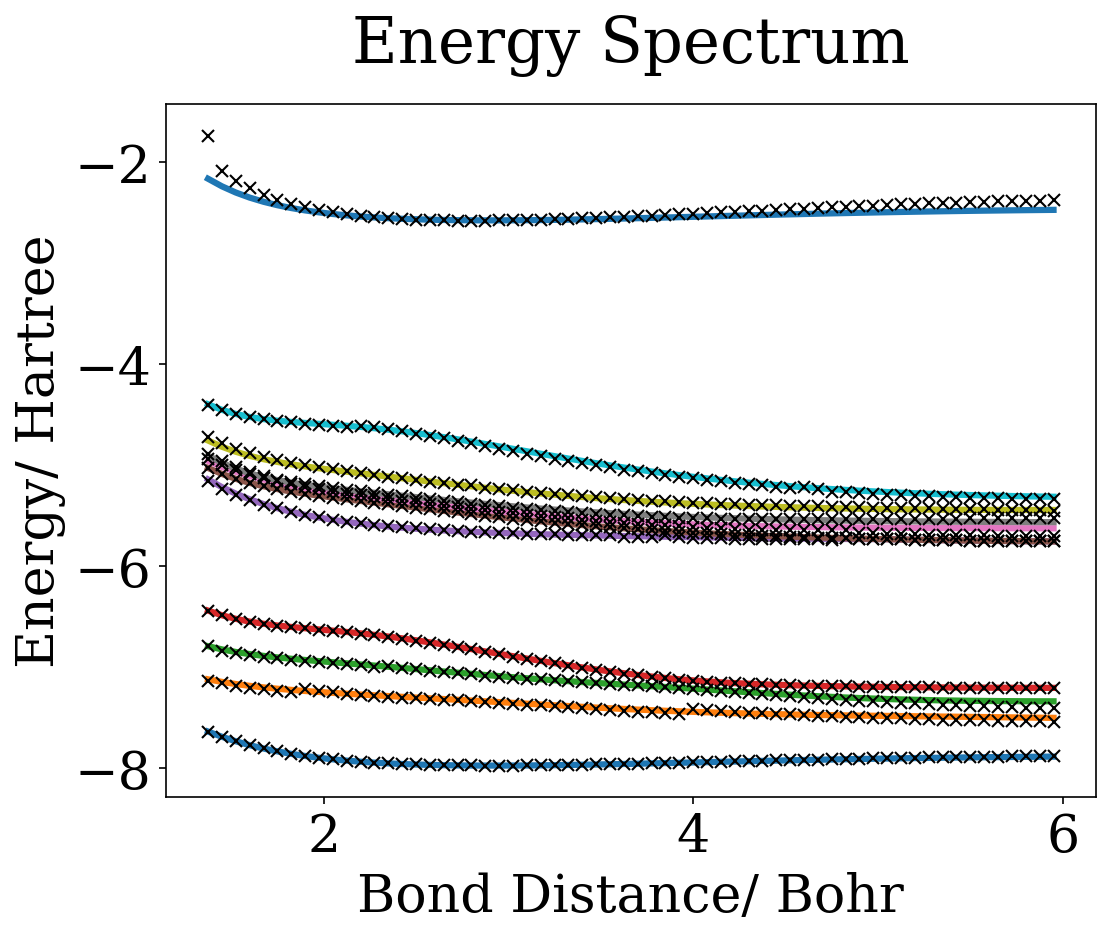}
    \caption{}
    \label{fig:CCD_LiH_acc}
    \end{subfigure}
    \hfill
    \begin{subfigure}[t]{0.43\textwidth}
    \includegraphics[width = \textwidth]{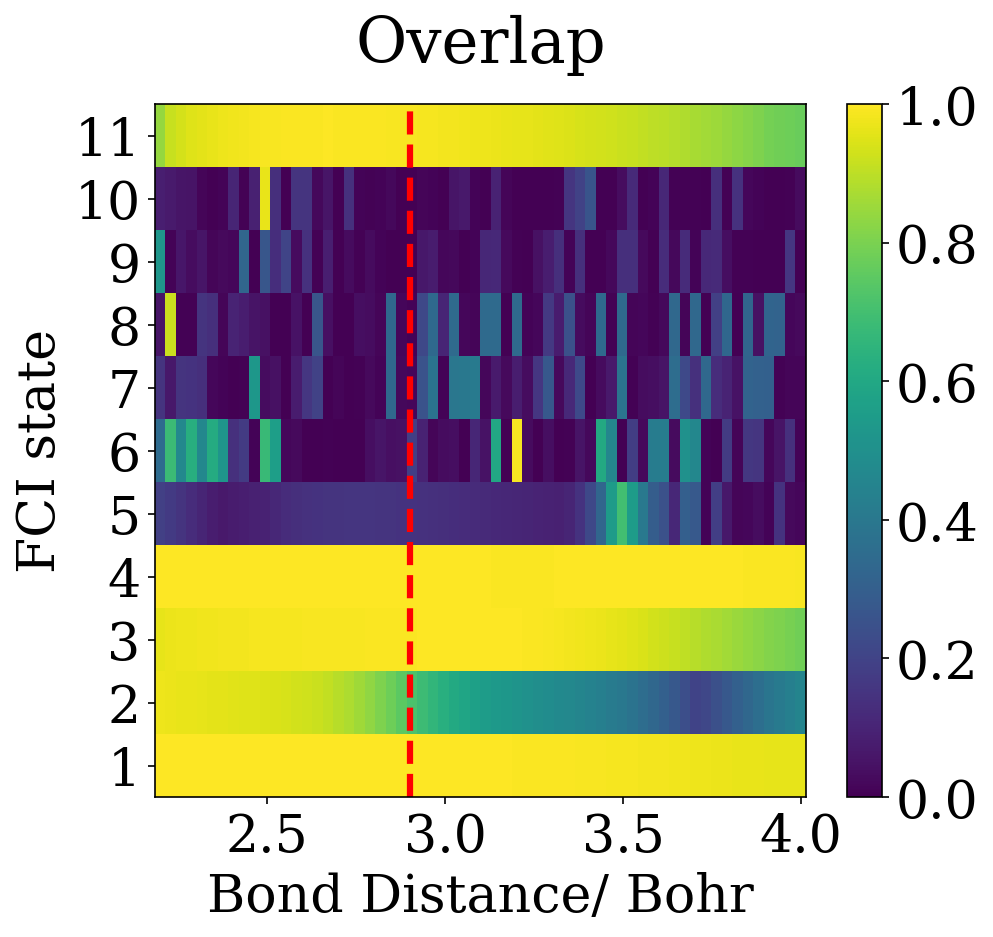}
    \caption{}
    \label{fig:CCD_LiH_ovlp}
    \end{subfigure}
    \caption{(a) PECs of LiH (solid lines) that can be accurately approximated by CCD energies (crosses). (b) Overlaps of the CCD states with the corresponding eigenstate. The dashed red line indicates the equilibrium geometry.}
\end{figure}

We initiate our energy and state approximation investigations by looking at the dissociation process of lithium hydride. We consider bond distances ranging from $1.375$ to $5.95$ bohr. Diagonalizing the Hamiltonian yields the full spectrum reported in Figure~\ref{fig:CCD_LiH_full} (see supplementary material) together with all roots to the CCD equations. Due to the large number of FCI solutions, we minimize the distance between FCI solutions and CCD solutions showing a much clearer picture of potentially well-resolved PECs, see Figure~\ref{fig:CCD_LiH_minima} in the supplementary material. This graph allows us to extract the PECs of the different states that are well-approximated using CCD, see Figure~\ref{fig:CCD_LiH_acc}. We find that 10 PECs are approximated up to $5\cdot 10^{-2}$ hartree for the entire dissociation process, however, we also note that for various bond distances, more CCD energies approximate FCI energies well, see Figure~\ref{fig:CCD_LiH_minima} and Table~\ref{tab:NumRootsLiH}. For the 10 PECs, we moreover compute the overlap between the CCD ket-states, i.e., $\ket{\Psi}
=
e^{\hat T} \ket{\Phi_0}$, and the approximated eigenstates, see Figure \ref{fig:CCD_LiH_ovlp}. We find that three of the four lowest energy states are well approximated both in terms of energy and states. The highest energy state stands out as well since it is well-approximated near the equilibrium, both in terms of energy as well as the eigenstate. However, as we transition away from the equilibrium the level of approximation deteriorates. A potential explanation for this observation is that the 1s orbital of the lithium atom forms a core making particle-hole excitations from the lowest orbital of LiH much less important than those from the valence $\sigma$ orbital. Thus, LiH is quite close to a two-electron problem for which CCSD is exact, and therewith, one could argue that the resulting near-exactness of CCSD for LiH means that the solutions of the CCSD equations (including wave functions) are expected to be numerically close to their full CI counterparts. We also note that a number of states show rather poor overlap with the corresponding eigenstates, which aligns with previous findings when investigating the root structure of H$_4$ clusters\cite{piecuch2000computational}. There are several potential reasons for this. First, the poor overlap may be detected when the eigenspace is higher dimensional, that is if the corresponding eigenvalue has multiplicity two or higher. Therefore some of the higher energy states reporting bad overlap might have CCD states that are close to the actual eigenspace -- or even within the eigenspace -- since we are only reporting overlap with one representative from the eigenspace. Second, it might be that the poor resolution via the ket-state is related to CC being a non-variational theory. In this case, the inclusion of a projection onto the bra-state, i.e., $e^{\hat T} \ket{\Phi_0} \bra{\Phi_0} (I + \hat \Lambda)e^{-\hat T}$ where $\hat \Lambda$ is a de-excitation operator defined via the dual variable of $t$ in the CC Lagrange formulation, could yield a more detailed insight. A thorough investigation of this is left for future work.

Investigating the root structure, we observe that the total number of CCD roots fluctuates along the bond stretching process, see Table~\ref{tab:NumRootsLiH}. This fluctuation could be caused by the continuous parameter curve of Hamiltonians crossing through the discriminant of the CC family, yielding a change in the root count. We find that any bound for a generic Hamiltonian severely overestimates the true number of roots for this system.

\begin{table}[h!]
    \centering
    \begin{tabular}{l|cccccccccc}
        Bond distances  & 1.375 & 1.6 & 1.9 & 2.2 & 2.5 & 3.1 & 4.0 & 4.9 & 5.8\\
        \hline
        $\#$ CCD sols. & 39 & 44 & 37 & 42 & 36 & 36 & 36 & 41 & 36\\
        $\#$ CCD real & 33 & 40 & 33 & 40 & 34 & 28 & 30 & 33 & 28\\
        $\#$ CCD approx.  & 22 & 25 & 20 & 23 & 18 & 19 & 17 & 18 & 14 
    \end{tabular}
    \caption{The number of CCD roots in LiH dissociation for selected bond distances. 
    By "$\#$ CCD approx."~we denote the energetically relevant CCD roots.
    For comparison, we recall that the B\'ezout bound led to $2^{36}$ solutions, the bound in Eq.~\eqref{eq:VarietyBound} led to 74 solutions, and CCdeg$_{4,8}$(\{2\}) is 73.}
    \label{tab:NumRootsLiH}
\end{table}

\subsection{Lithium hydride -- CCSD}

\begin{figure}[h!]
    \centering
    \begin{subfigure}[t]{0.45\textwidth}
    \includegraphics[width =\textwidth]{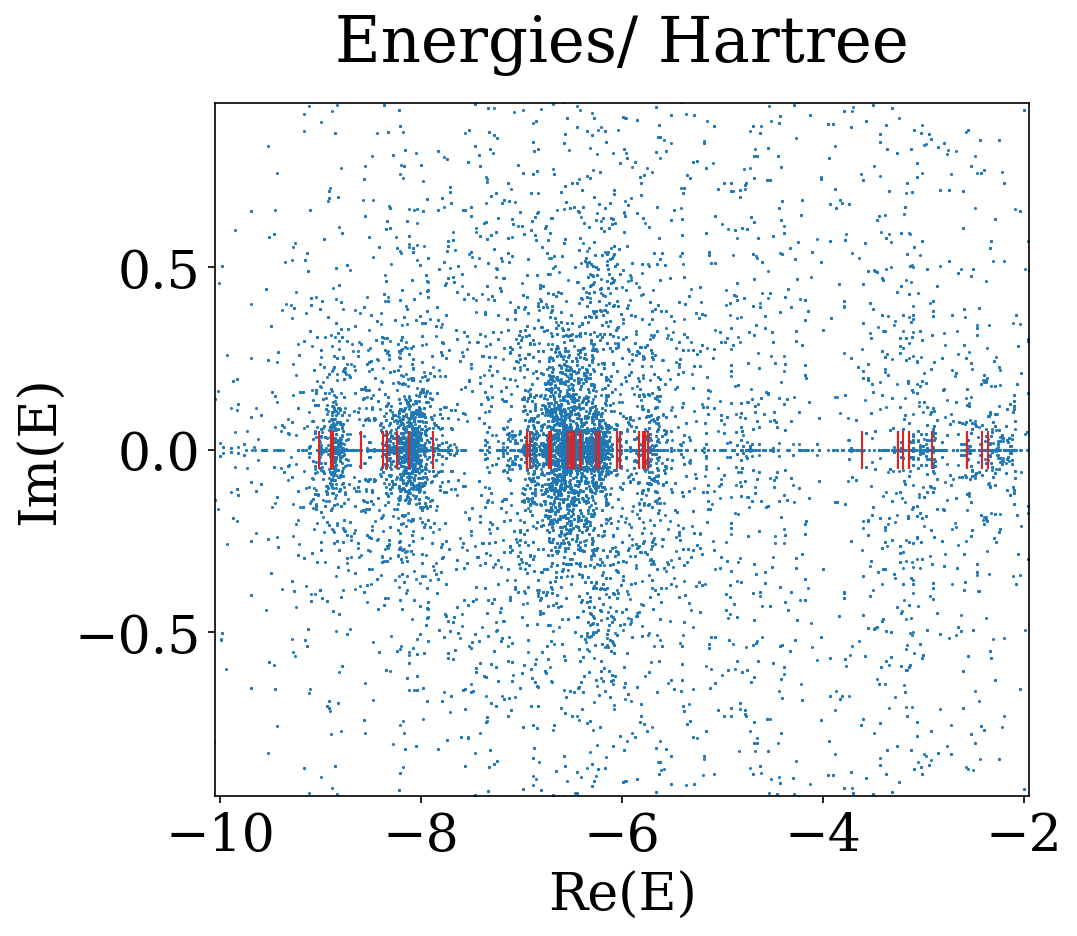}
    \caption{}
    \label{fig:CCSD_LiH_all}
    \end{subfigure}
    \hfill
    \begin{subfigure}[t]{0.45\textwidth}
    \includegraphics[width = \textwidth]{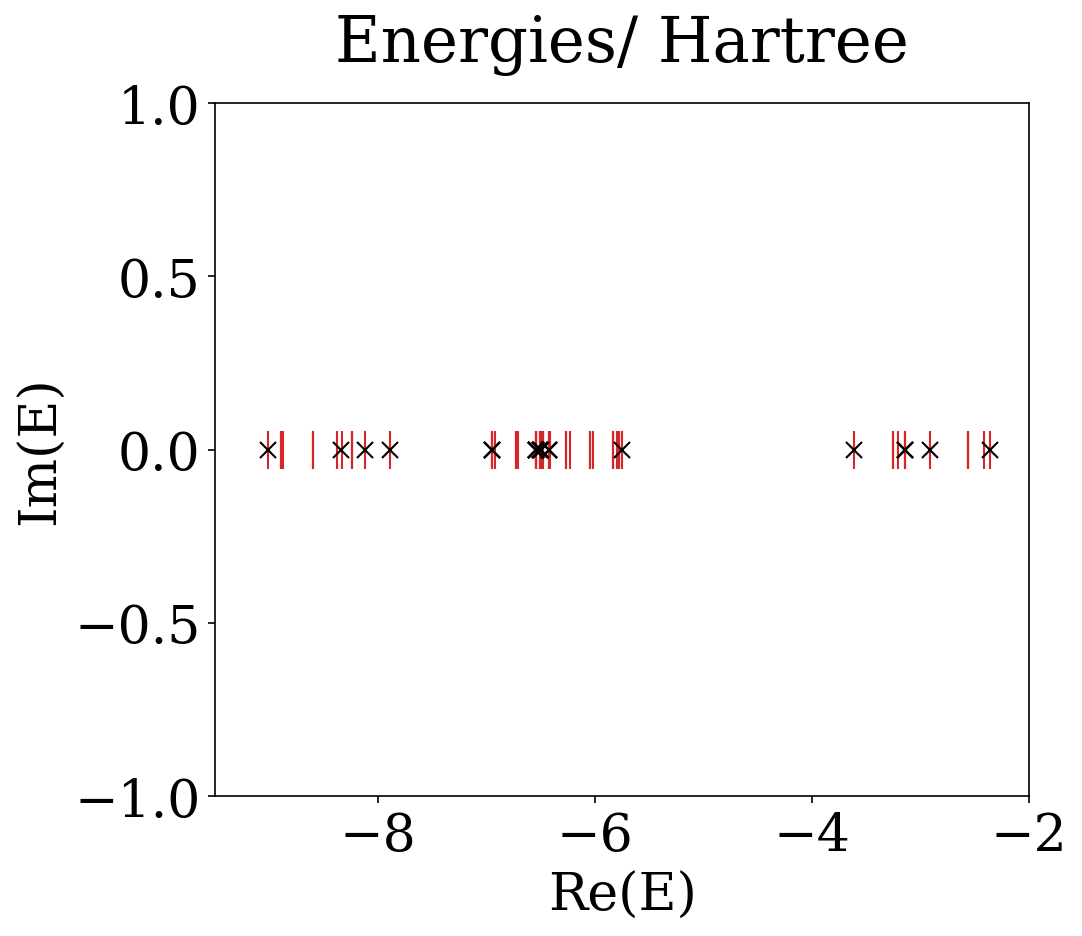}
    \caption{}
    \label{fig:CCSD_LiH_kept}
    \end{subfigure}
    \caption{(a) FCI solutions (red lines) together with all CCSD energies (b) CCSD energies that approximate FCI energies up to $ 10^{-3}$ hartree.}
\end{figure}

We moreover perform CCSD computations on lithium hydride at the bond distance 2.875 bohr -- close to the molecule's equilibrium. Computing all roots to CCSD equations yields $15\,954$ roots, $2\,170$ of which yield real-values energies, and $1\,280$ are real-valued solutions. 
Visualizing all $15\,954$ CCSD energies shows that the energies tend to cluster around the FCI solutions, see Figure~\ref{fig:CCSD_LiH_all}. Being able to compare with the true Hamiltonian spectrum reveals that only 26 of these CCSD solutions yield energies that are close to an FCI energy up to $10^{-3}$ hartree, see Figure~\ref{fig:CCSD_LiH_kept}.

\begin{figure}[h!]
\includegraphics[width = 0.45\textwidth]{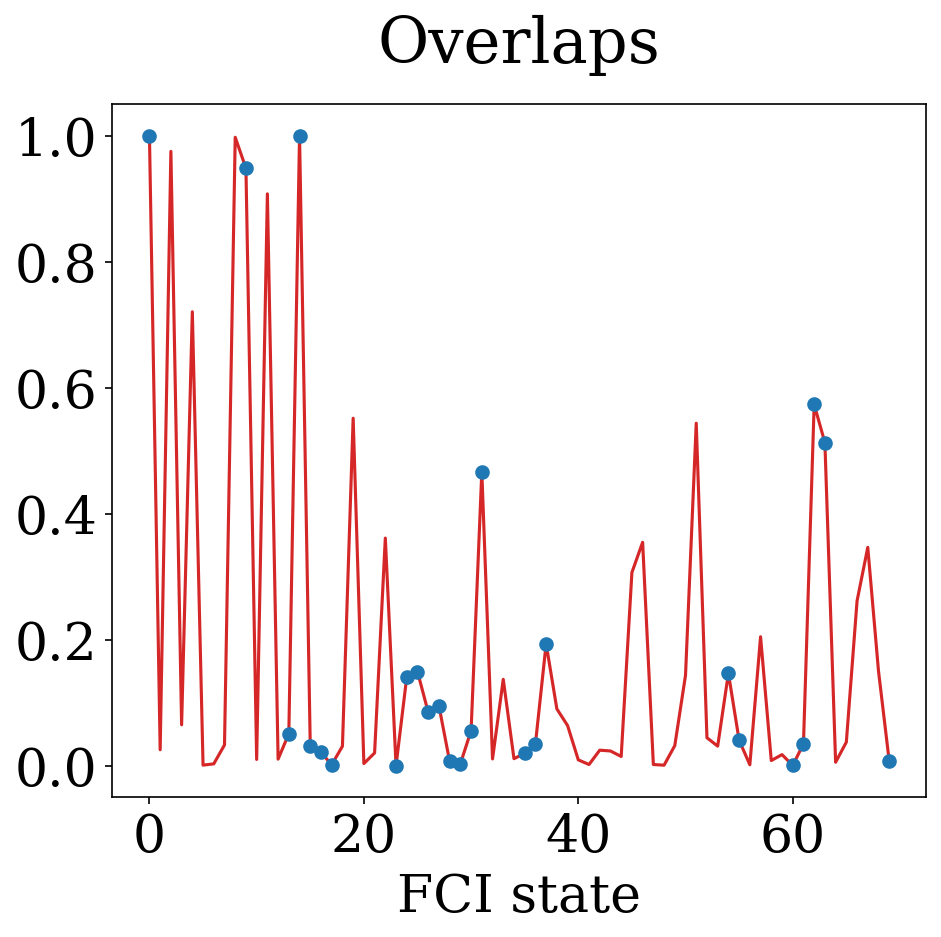}
\caption{Overlap of the energetically closest CCSD state with the corresponding eigenstate. 
The blue dots show the 26 energetically relevant CCSD states.}
\label{fig:LiHOverlap}
\end{figure}

For all 70 eigenstates, we compute the energetically closest CCSD state together with the overlap with the corresponding states, see Figure~\ref{fig:LiHOverlap}. We find that three of the 26 energetically relevant CCSD states have good overlaps with the targeted states. Therefore there are at least three CCSD states that are well approximated both in terms of energy and state. 

\subsection{Dissociating systems of hydrogen}

We now investigate the dissociation of (H$_2$)$_2$ planar model systems in different geometries~\cite{paldus1993application, jankowski1980applicability,piecuch2000computational}.

\subsubsection{(H$_2$)$_2$ dissociation (D$_{2{\rm h}}$ symmetry)}

\begin{figure}[h!]
\includegraphics[width = 0.3\textwidth]{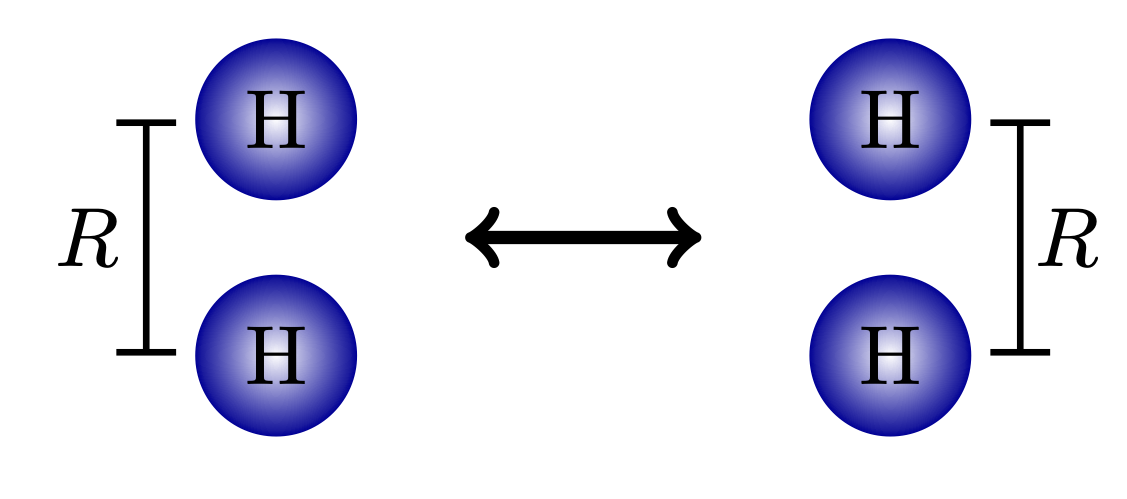}
\caption{Schematic depiction of the dissociation process of (H$_2$)$_2$ in D$_{2 {\rm h}}$ configuration. The parameter $R = 1.4$ bohr. }
\label{fig:H4D2h}
\end{figure}

The bond stretching procedure for (H$_2$)$_2$ in D$_{2{\rm h}}$ symmetry is sketched in Figure~\ref{fig:H4D2h}. We consider an intra-molecular bond distance of $R = 1.4$ bohr, which corresponds to the equilibrium geometry of H$_2$; this is kept fixed during the dissociation process. The inter-molecular distance is varied from 1.5 to 4.0 bohr.

\begin{figure}[h!]
    \centering
    \begin{subfigure}[t]{0.45\textwidth}
    \includegraphics[width =\textwidth]{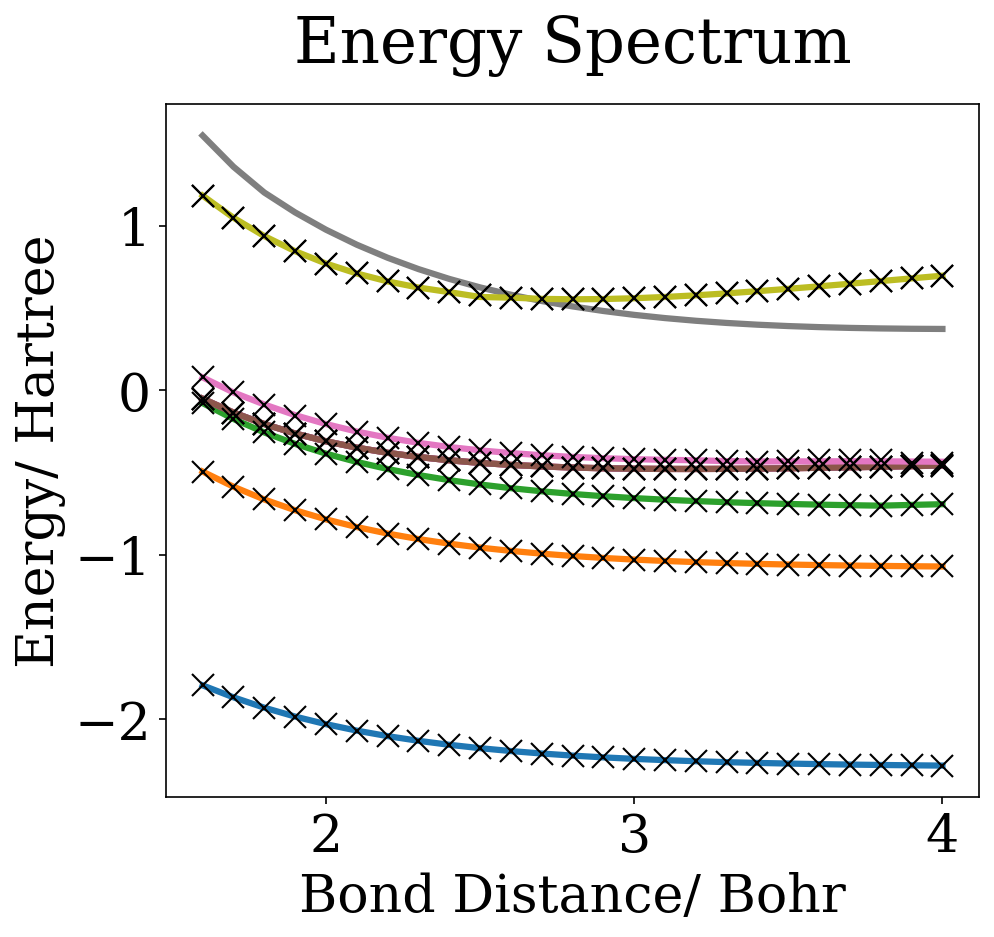}
    \caption{}
    \label{fig:CCD_h4_D2h_semi}
    \end{subfigure}
    \hfill
    \begin{subfigure}[t]{0.43\textwidth}
    \includegraphics[width = \textwidth]{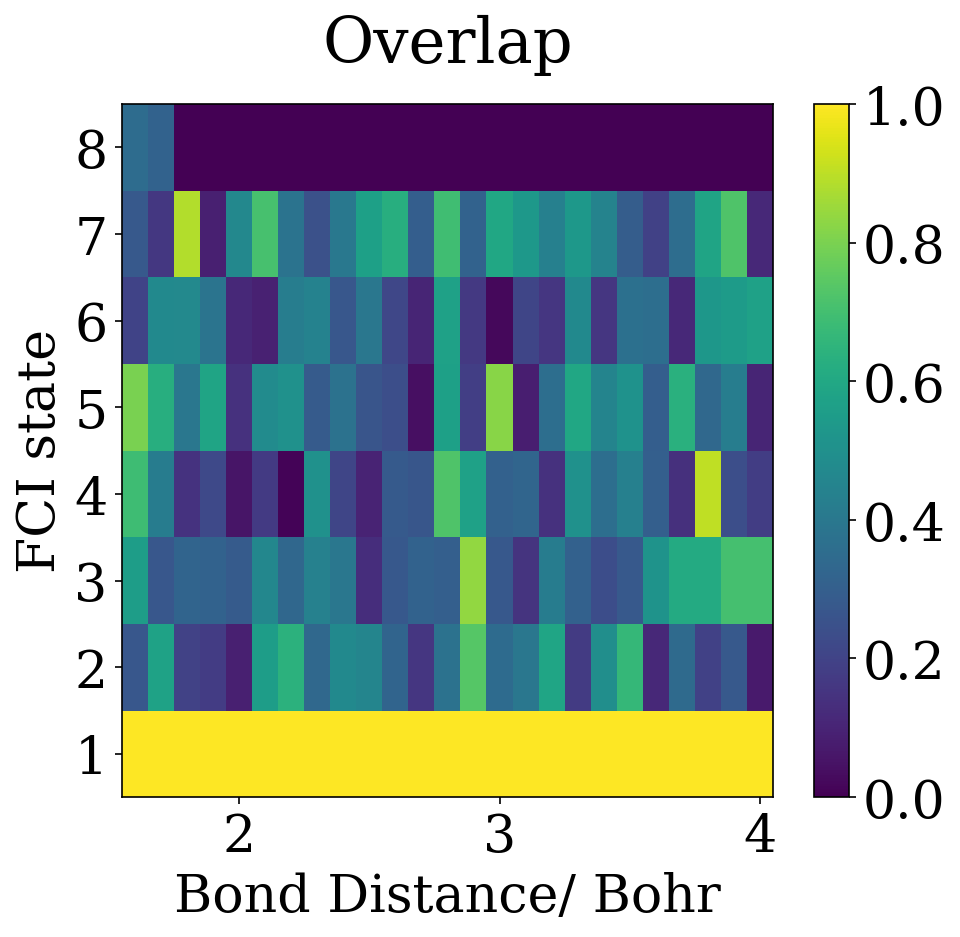}
    \caption{}
    \label{fig:CCD_h4_D2h_ovlp}
    \end{subfigure}
    \caption{(a) PECs of (H$_2$)$_2$ in D$_{2{\rm h}}$ symmetry (solid lines) that are accurately described by CCD energies (crosses). (b) Overlap of the CCD states with the corresponding eigenstate.}
\end{figure}

The full spectrum together with the CCD solutions can be found in Figure~\ref{fig:CCD_h4_D4_full} in the Supplementary material, and the PECs together with the closest CCD solutions are reported in Figure~\ref{fig:CCD_D4_min}. We extract the PECs that are well-approximated using CCD and report them in Figure~\ref{fig:CCD_h4_D2h_semi}. Here, we observe that eight CCD energies approximate FCI PECs well. For these eight CCD states, we compute the overlap with the corresponding eigenstate, see Figure~\ref{fig:CCD_h4_D2h_ovlp}. 
We see that the ground state is well approximated, both in terms of energy as well as the state. The following six states (i.e.~states two to seven in Figure~\ref{fig:CCD_h4_D2h_ovlp}) have intermediate overlaps with the targeted states; the last state has a poor overlap with the targeted state. 

In Table~\ref{tab:NumRootsH4D2h} we report the number of CCD roots yielding real-valued energies and energetic relevant CCD roots for selected bond distances. We observe that the number of CCD roots fluctuates along the bond stretching procedure. 

\begin{table}[h!]
    \centering
    \begin{tabular}{l|cccccc}
        Bond distances & 1.6 & 1.9 & 2.1 & 2.5 & 3.0 & 3.9 \\
        \hline
        $\#$ CCD sols. & 72 & 53 & 68 & 72 & 63 & 70\\
        $\#$ CCD real & 66 & 45 & 60 & 64 & 55 & 64\\
        $\#$ CCD approx.  & 32 & 26 & 33 & 32 & 33 & 34
    \end{tabular}
    \caption{The number of roots along the (H$_2$)$_2$ dissociation in D$_{2 {\rm h}}$ configuration for selected bond distances. 
    By "$\#$ CCD approx."~we denote the energetically relevant CCD roots.
    For comparison, we recall that the B\'ezout bound is $2^{36}$, the bound in Eq.~\eqref{eq:VarietyBound} yields 74, and CCdeg$_{4,8}$(\{2\}) is 73.}
    \label{tab:NumRootsH4D2h}
\end{table}

\subsubsection{(H$_2$)$_2$ dissociation (D$_{\infty {\rm h}}$ symmetry)}

\begin{figure}[h!]
\includegraphics[width = 0.3\textwidth]{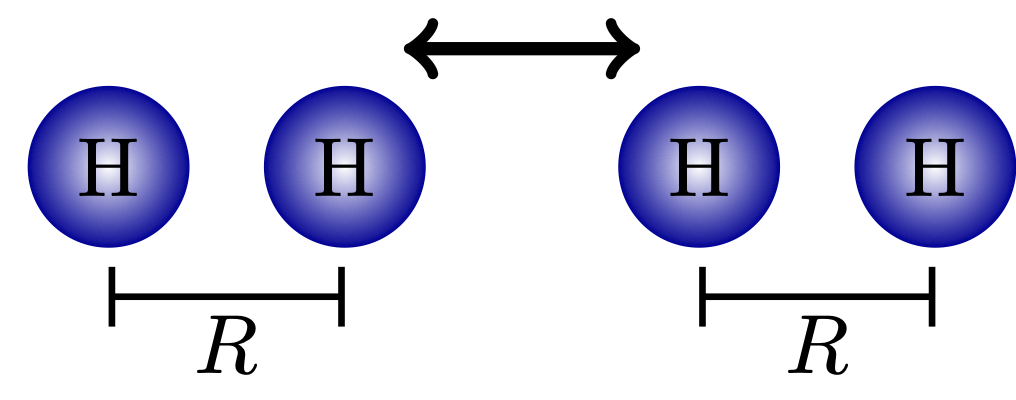}
\caption{Schematic depiction of the dissociation process of (H$_2$)$_2$ in D$_{\infty {\rm h}}$ configuration. The parameter $R = 1.4$ bohr. }
\label{fig:H4Dinfh}
\end{figure}

Next, we investigate the (H$_2$)$_2$ dissociation in D$_{\infty {\rm h}}$ symmetry, see Figure~\ref{fig:H4Dinfh} for a sketch of the dissociation process. 
The intra-molecular distance is again set to $R = 1.4$ bohr and is kept fixed during the dissociation process. The inter-molecular distance is varied from 1.5 to 4.0 bohr.

\begin{figure}[h!]
    \centering
    \begin{subfigure}[t]{0.45\textwidth}
    \includegraphics[width =\textwidth]{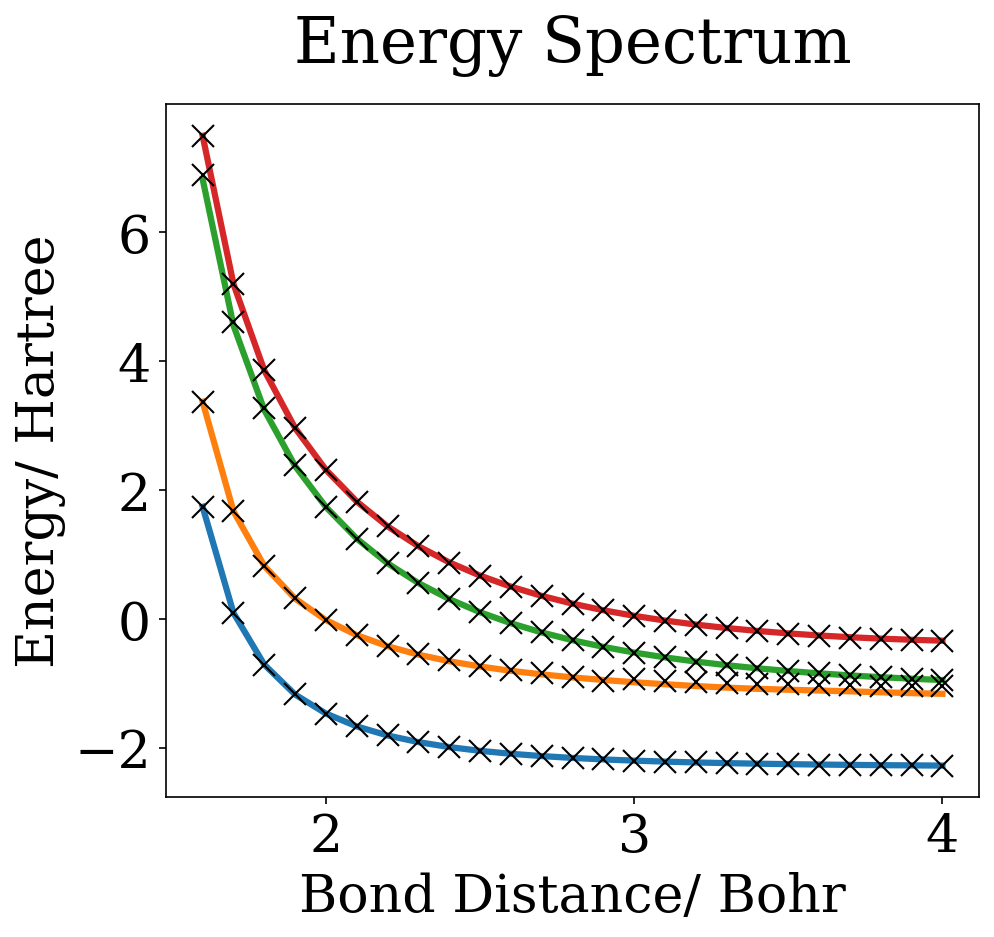}
    \caption{}
    \label{fig:CCD_h4_Dinfh_semi}
    \end{subfigure}
    \hfill
    \begin{subfigure}[t]{0.43\textwidth}
    \includegraphics[width = \textwidth]{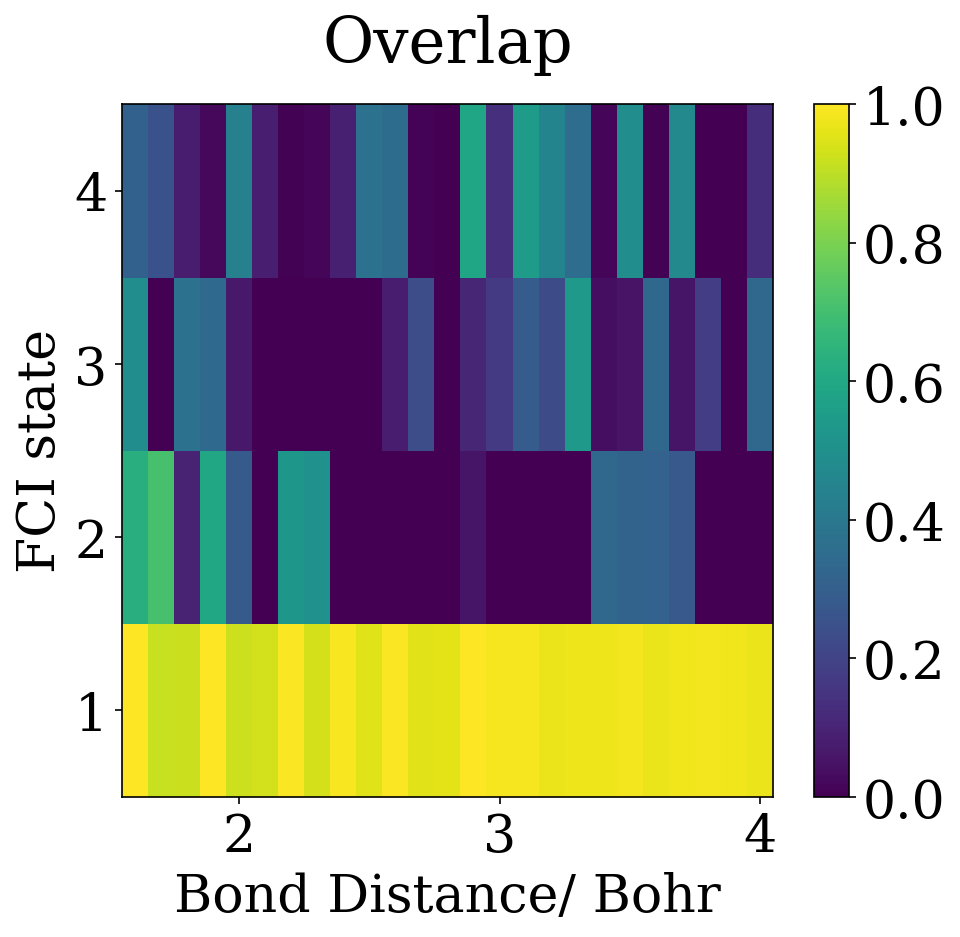}
    \caption{}
    \label{fig:CCD_h4_D2h_olvp}
    \end{subfigure}
    \caption{(a) PECs of (H$_2$)$_2$ in D$_{\infty{\rm h}}$ symmetry (solid lines) that are accurately described by CCD energies (crosses). (b)  Overlap of the CCD states with the corresponding eigenstate.}
\end{figure}

The full spectrum and CCD solutions is presented in Figure~\ref{fig:CCD_h4_P4_full} in the Supplementary material, with the PECs and their closest CCD solutions detailed in Figure~\ref{fig:CCD_h4_P4_min}. We have identified PECs that are accurately approximated by CCD, as shown in Figure~\ref{fig:CCD_h4_Dinfh_semi}, where four CCD energies closely match the FCI PECs. The overlap of these four CCD states with their respective eigenstates is analyzed in Figure~\ref{fig:CCD_h4_D2h_olvp}. Consistent with observations from the (H$_2$)$_2$ in D$_{2{\rm h}}$ configuration, the ground state demonstrates a high degree of accuracy, while the other three states exhibit intermediate to poor overlap with their targeted eigenstates.

Investigating the root structure, we observe that -- similar to (H$_2$)$_2$ in D$_{2{\rm h}}$ symmetry -- the total number of CCD roots fluctuates along the bond stretching procedure, see Table~\ref{tab:NumRootsH4Dinfh}. Note that for (H$_2$)$_2$ in D$_{\infty {\rm h}}$ symmetry, the total number of CCD roots is lower than the number of CCD roots (H$_2$)$_2$ in D$_{2{\rm h}}$ symmetry. In particular, any bound for a generic Hamiltonian severely overestimates the true number of roots for this system. 

\begin{table}[h!]
    \centering
    \begin{tabular}{l|ccccccc}
        Bond distances & 1.6 & 1.9 & 2.1 & 2.5 & 3.0 & 3.9\\
        \hline
        $\#$ CCD sols. & 48 & 44 & 44 & 44 & 45 & 39\\
        $\#$ CCD real & 46 & 42 & 40 & 38 & 37 & 35\\
        $\#$ CCD approx.  & 13 & 19 & 12 & 14 & 16 & 21 \\
    \end{tabular}
    \caption{The number of roots along the (H$_2$)$_2$ dissociation in D$_{\infty {\rm h}}$ configuration for selected bond distances. 
    By "$\#$ CCD approx."~we denote the energetically relevant CCD roots.
    For comparison, we recall that the B\'ezout bound led to $2^{36}$ solutions, the bound in Eq.~\eqref{eq:VarietyBound} led to 74 solutions, and CCdeg$_{4,8}$(\{2\}) is 73.}
    \label{tab:NumRootsH4Dinfh}
\end{table}

\subsection{H$_4$ disturbed on a circle}

\begin{figure}[h!]
    \centering
    \includegraphics[width = 0.45\textwidth]{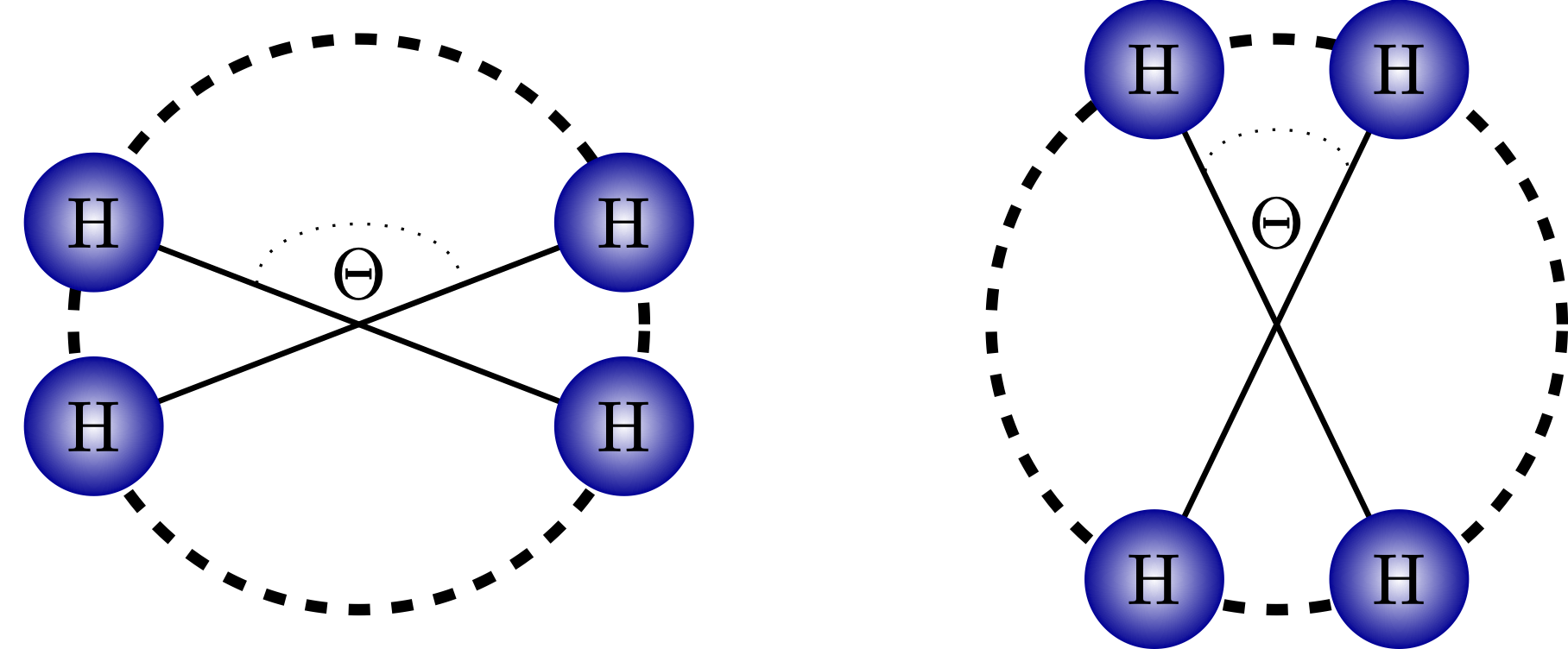}
    \caption{Schematic depiction of the H$_4$ model undergoing a symmetric disturbance on a circle modeled by the angle~$\Theta$.}
    \label{fig:H4Circle}
\end{figure}

We proceed by investigating a variant of the H$_4$ model consisting of four hydrogen atoms symmetrically distributed on a circle of radius $R = \sqrt{2}$~bohr~\cite{bulik2015can, Loos2021, kossoski2021excited}. We here investigate different geometries defined by the parameter $\Theta \in (0^\circ, 180^\circ)$ which describes the angle formed by the line segments connecting two hydrogen atoms that are placed point symmetrically with respect to the origin of the circle. For a schematic depiction of this process see Fig.~\ref{fig:H4Circle}.

\begin{figure}[h!]
    \centering
    \begin{subfigure}[t]{0.45\textwidth}
    \includegraphics[width =\textwidth]{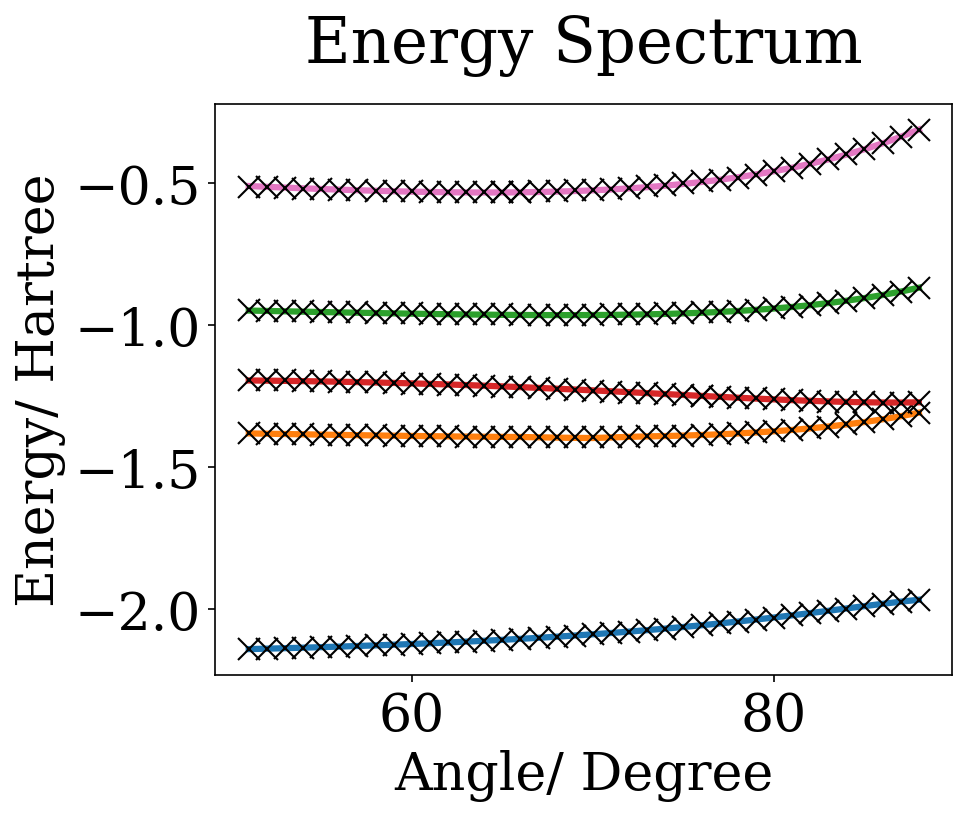}
    \caption{}
    \label{fig:CCD_h4_P4D4_semi}
    \end{subfigure}
    \hfill
    \begin{subfigure}[t]{0.40\textwidth}
    \includegraphics[width = \textwidth]{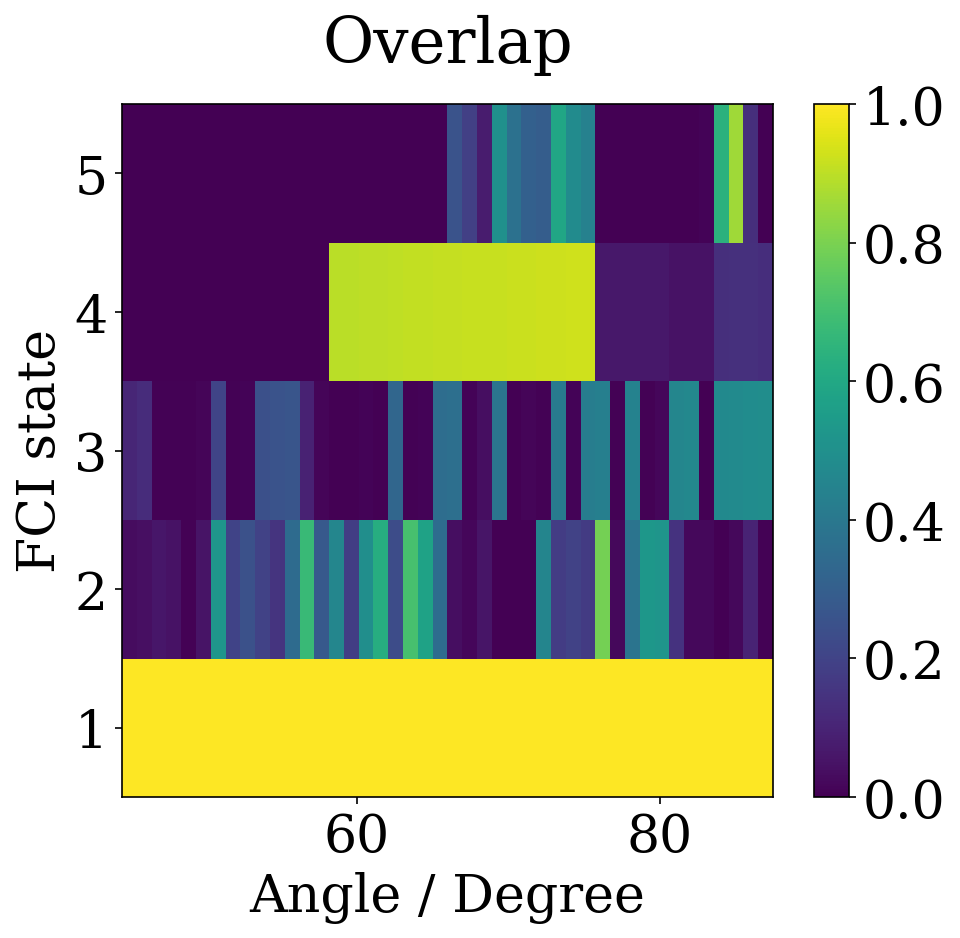}
    \caption{}
    \label{fig:CCD_h4_P4D4_ovlp}
    \end{subfigure}
    \caption{(a) PECs of H$_4$ symmetrically distributed on a cycle (solid lines) that can be accurately reconstructed by CCD roots (crosses). (b) Overlap of the CCD states with the corresponding eigenstate.}
\end{figure}

The full spectrum together with the CCD solutions can be found in Figure~\ref{fig:CCD_h4_P4D4_full} in the Supplementary material, and the PECs together with the closest CCD solutions are reported in Figure~\ref{fig:CCD_h4_P4D4_min}. We extract the PECs that are well-approximated using CCD which are reported in Figure~\ref{fig:CCD_h4_P4D4_semi}. Note that it is well-known that for this system CCD does not perform well in the region close to 90$^\circ$, due to strong degeneracies~\cite{paldus1993application}. This results in a dramatic reduction of the total number of roots, we find three roots at $90^\circ$ and six roots at $89^\circ$. For completion, these energies are included in the supplementary material, however, since the focus of this work is on the root structure and potentially physically relevant roots accessible by single-reference CC theory, we focus on cases where single-reference CC theory provides reasonable approximations and therefore excluded these points in Figure~\ref{fig:CCD_h4_P4D4_semi}. Outside of this challenging region, we find five solutions that approximate PECs well. For these five solutions, we compute the overlap with the corresponding eigenstate, see Figure~\ref{fig:CCD_h4_P4D4_ovlp}. 
Consistent with the previous H$_4$ systems, this model system shows the effect of CCD states accurately resolving excited energies but providing poor overlap to the exact eigenstates. The only state that shows a good approximation to its eigenstate is the CCD ground state.

Similar to (H$_2$)$_2$ in D$_{\infty {\rm h}}$ symmetry, we observe that generic bounds severely overestimate the number of CCD roots, though this effect is further amplified for H$_4$ symmetrically distributed on a circle, see Table~\ref{tab:NumRootsH4Circle}.

\begin{table}[h!]
    \centering
    \begin{tabular}{l|ccccccc}
        Angles in degrees & 45 & 55 & 65 & 75 & 85\\
        \hline
        $\#$ CCD sols. & 43 & 44 & 43 & 44 & 36 \\
        $\#$ CCD real & 35 & 36 & 35 & 36 & 28 \\
        $\#$ CCD approx.  & 18 & 18 & 19 & 21 & 22  
    \end{tabular}
    \caption{The number of roots of H$_4$ symmetrically distributed on a circle for selected bond distances. By "$\#$ CCD approx."~we denote the energetically relevant CCD roots.
    For comparison, we recall that the B\'ezout bound led to $2^{36}$ solutions, the bound in Eq.~\eqref{eq:VarietyBound} led to 74 solutions, and CCdeg$_{4,8}$(\{2\}) is 73.}
    \label{tab:NumRootsH4Circle}
\end{table}

\section{Conclusion}

The introduction highlighted critical unanswered questions in the study of the coupled cluster (CC) equations' algebraic structures, including the total number of solutions, the best approximation element, and the feasibility of approximating excited states with higher-order roots of single reference CC equations. The ability to compute the full solution spectrum to the CC equations enabled us to tackle these questions for concrete physical systems.

Most remarkably, our results demonstrate that for lithium hydride in multiple geometries, higher-order CC solutions not only provide high-accuracy approximations to the energies but also to the states themselves. For a broader range of systems including multiple (H$_2$)$_2$ planar model systems in different geometries, we found that high-order CC solutions accurately predict several full potential energy curves, though they possibly provide a poor approximation to the eigenstates. While this effect has been previously reported, it requires further investigation to fully understand it and ultimately leverage it for practical applications.

Additionally, we examined the total number of roots and the effectiveness of various bounds developed for generic Hamiltonians. Our simulations indicate that while the most recent bounds perform exceptionally well for generic systems, they significantly overestimate the number of solutions in physical systems. More precisely, for all considered systems, we computed between $30$-$60$ real solutions of which only $10$-$30$ approximate physical states, compared to CC degree of $72$. This excess of solutions gets amplified for CCSD: Solving the CC equations of a generic Hamiltonian yields $16\,952\,996$ solutions, compared to $1280$ solutions that yield real-valued energies of the CC equations describing LiH. This overestimation by a factor of over ten thousand indicates that to further improve the bounds of the CC equations, specific structures of the Hamiltonian have to be taken into account. This includes e.g. a spin-integrated or spin-adapted description making the system of polynomial equations more physical and closer to conventional computational quantum chemistry formulations.

\begin{acknowledgement}

The authors thank Piotr Piecuch for insightful discussions. Additionally, we thank the anonymous reviewers for their comments, which have significantly contributed to the improvement of this manuscript.

\end{acknowledgement}

\section{Associated Content}

Supporting Information Available:


We present graphical representations of the full energy spectra alongside the real-valued CCD energies for several Hamiltonians: (H$_2$)$_2$ in both D$_{2{\rm h}}$ and D$_{\infty {\rm h}}$ configurations, H$_4$ symmetrically distorted on a circle, and lithium hydride. Additionally, we analyze the CCD energies by minimizing the difference between these energies and the spectral values of the Hamiltonians. This yields the CCD energies that best approximate the spectral values of the Hamiltonians.

\noindent
This information is available free of charge at the website: http://pubs.acs.org/

\bibliography{lib.bib}



\section*{Supplementary material}

# Supporting Information Available

## Dissociating systems of hydrogen

We begin by reporting the full energy spectrum of the Hamiltonian describing $(H_2)_2$ in $D_{2h}$ symmetry together with all CCD roots, see Figure 13a. Given the large number of solutions, we minimize over the difference between spectral values of the Hamiltonian and the found CCD energies. This provides further insight into energies that are well approximated by CCD, see Figure 13b.

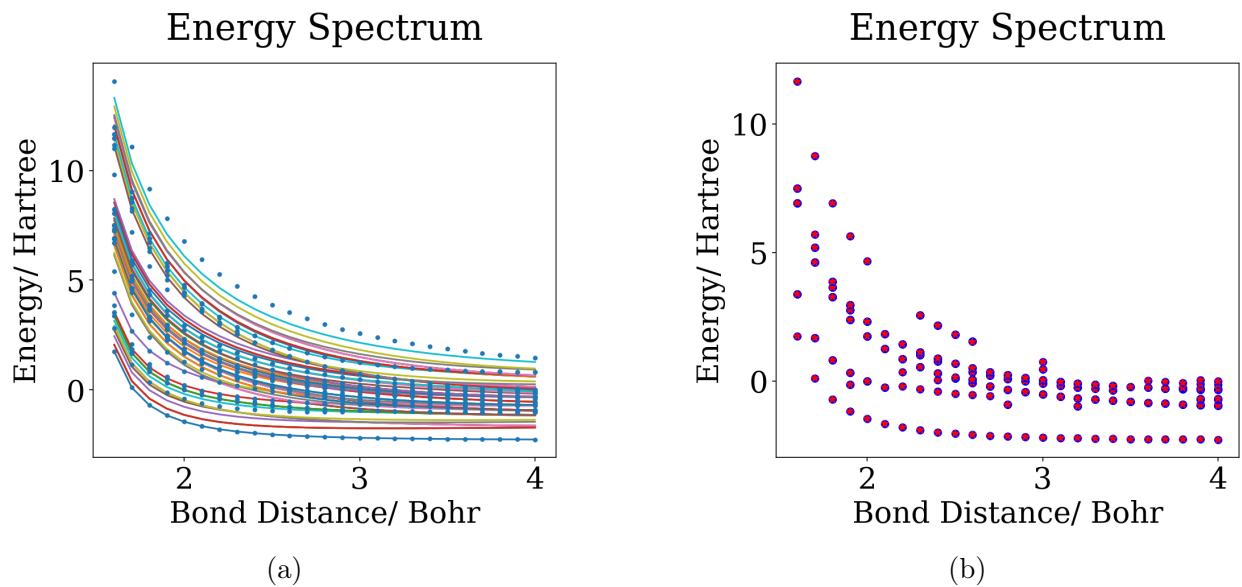

Figure 13: (a) Full spectrum (b) Minimal difference.

Similarly, we proceed for $(H_2)_2$ in $D_{\infty h}$ symmetry. We reporting the full energy spectrum of the Hamiltonian describing $(H_2)_2$ in $D_{\infty h}$ symmetry together with all CCD roots, see Figure 14a Given the large number of solutions, we minimize over the difference between spectral values of the Hamiltonian and the found CCD energies. This provides further insight into energies that are well approximated by CCD, see Figure 14b.

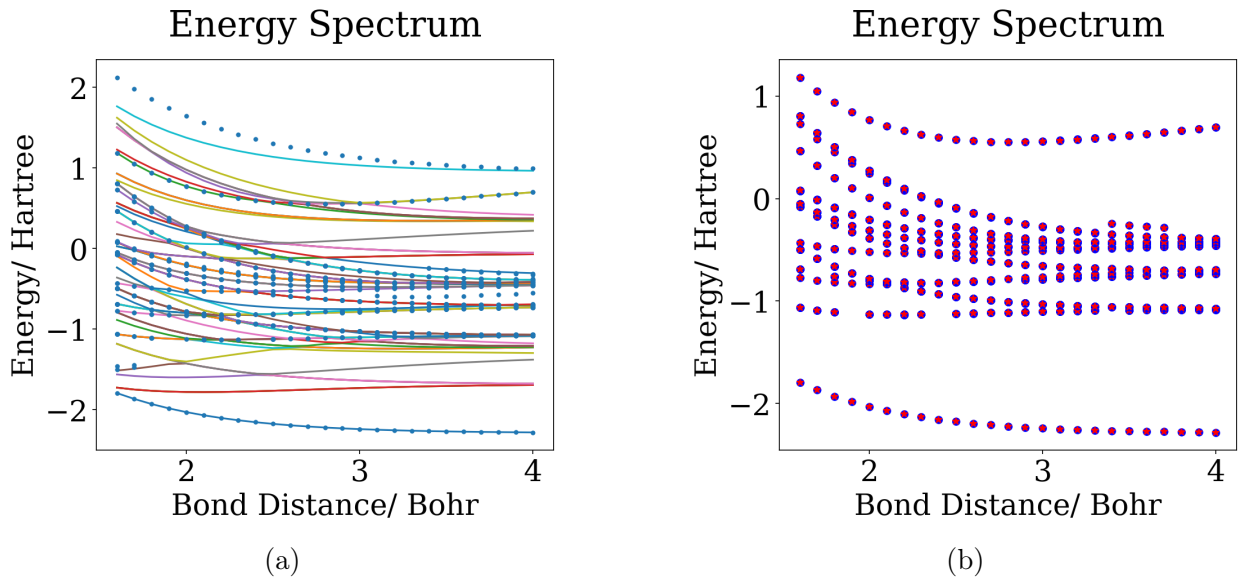

Figure 14: (a) Full spectrum (b) Minimal difference.

# H$_4$ disturbed on a circle

Similarly, we proceed for H$_4$ disturbed on a circle. We report the full energy spectrum of the Hamiltonian describing H$_4$ disturbed on a circle together with all CCD roots, see Figure 15a. Given the large number of solutions, we minimize over the difference between spectral values of the Hamiltonian and the found CCD energies. This provides further insight into energies that are well approximated by CCD, see Figure 15b.

# Lithium hydride

We report the full energy spectrum of the Hamiltonian describing LiH together with all CCD roots, see Figure 16a Given the large number of solutions, we minimize over the difference between spectral values of the Hamiltonian and the found CCD energies. This provides further insight into energies that are well approximated by CCD, see Figure 14b.

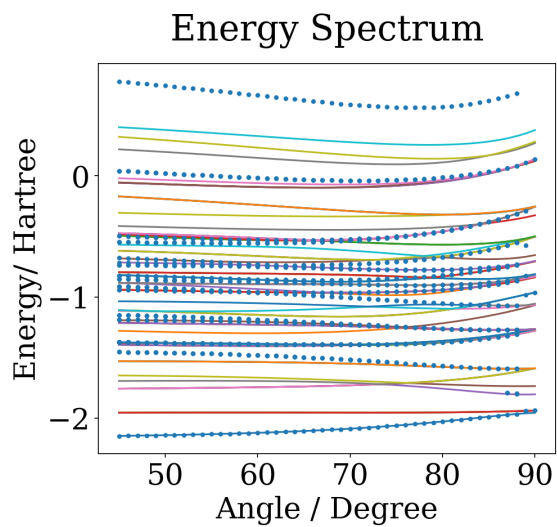
(a)

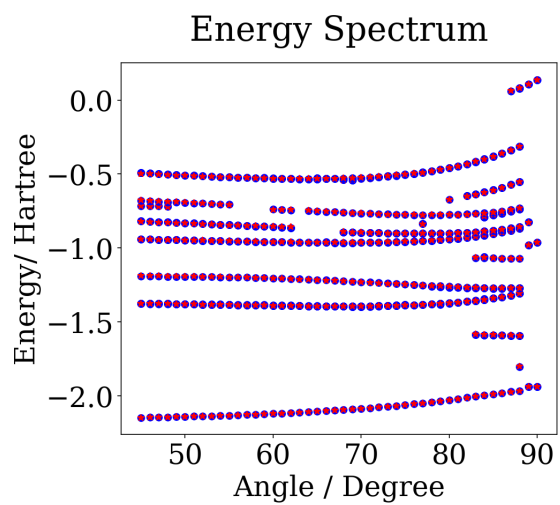
(b)

Figure 15: (a) Full spectrum (b) Minimal difference.

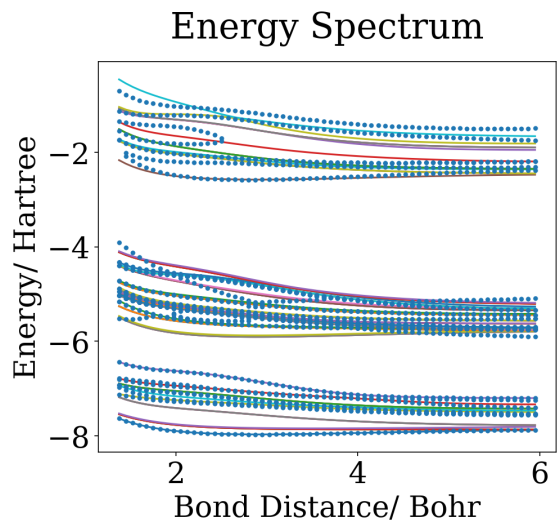
(a)

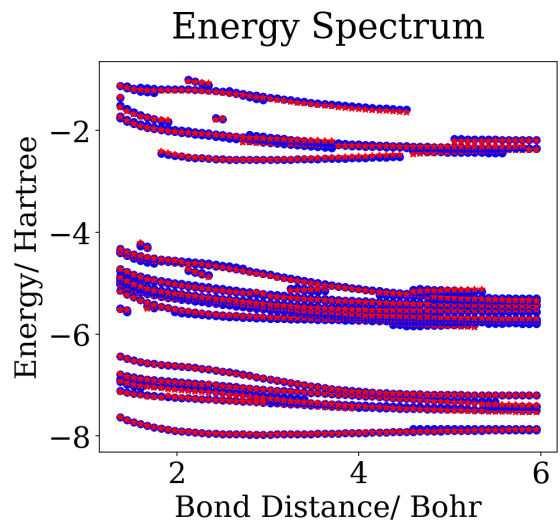
(b)

Figure 16: (a) Full spectrum (b) Minimal difference.

# TOC Graphic

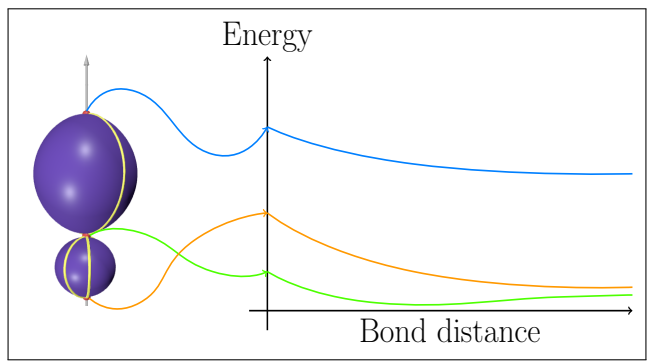



\end{document}